\documentclass{article}
\usepackage{amsmath,amsfonts,amsthm}
\usepackage{mathrsfs} 
\usepackage{authblk}
\usepackage{xcolor}
\usepackage{IEEEtrantools}
\usepackage{mathtools}
\usepackage{bm}
\usepackage[version=3]{mhchem}
\newcommand{\eps}{\varepsilon}
\newcommand{\RR}{\mathbb{R}}
\newcommand{\ud}{\,\mathrm{d}}
\newcommand{\weak}[1]{\stackrel{#1}{\rightharpoonup}}
\usepackage{hyperref}

\newtheorem{definition}{Definition}
\newtheorem{theorem}{Theorem}
\newtheorem{proposition}{Proposition}
\newtheorem{remark}{Remark}

\newcommand{\eqlabel}[1]{\addtocounter{equation}{-1}\refstepcounter{equation}\label{#1}}

\title{Second-order asymptotic expansion and thermodynamic interpretation of a fast-slow Hamiltonian system}

\begin{document}

\author[$\dagger$]{Matthias Klar}

\author[$\ddag$]{Karsten Matthies}

\author[$\dagger$]{Johannes Zimmer}

\affil[$\dagger$] {Technische Universität München, School of Computation, Information and Technology,
  Department of Mathematics, 85748 Garching, Germany}
  
\affil[$\ddag$]{Department of Mathematical Sciences, University of Bath, Bath BA2 7AY, UK}


\maketitle

\begin{abstract}This article includes a short survey of selected averaging and dimension reduction techniques for deterministic fast-slow systems. This survey includes, among others, classical techniques, such as the WKB approximation or the averaging method, as well as modern techniques, such as the GENERIC formalism. The main part of this article combines ideas of some of these techniques and addresses the problem of deriving a reduced system for the slow degrees of freedom (DOF) of a fast-slow Hamiltonian system. In the first part, we derive an asymptotic expansion of the averaged evolution of the fast-slow system up to second-order, using weak convergence techniques and two-scale convergence. In the second part, we determine quantities which can be interpreted as temperature and entropy of the system and expand these quantities up to second-order, using results from the first part. The results give new insights into the thermodynamic interpretation of the fast-slow system at different scales.~\footnote{This article belongs to the themed collection: Mathematical Physics and Numerical Simulations of Many-Particle Systems; V. Bach and L. Delle Site (eds.)}
\end{abstract}

\section{Introduction}
A common strategy for analysing how physical properties of macroscopic objects change over time is to model
the system on the microscopic scale and then derive macroscopic information, either through analytical
considerations or computer simulations.

This method has many advantages. It is conceptually easy to develop a model on the microscopic scale that
describes a macroscopic object to high accuracy. This model can then be analysed under different
circumstances, either analytically or with computer simulations, often under conditions that allow insight
into the object's properties which cannot be derived based on real world experimental data alone. Moreover,
modelling the physical properties of a macroscopic object based on well established physical laws on the
microscopic scale allows to control and analyse its most fundamental components, which are often more
mathematically sound than phenomenologically derived laws on the macroscopic scale. Finally, provided the
model on the microscopic scale reflects the actual physical properties to high accuracy, the derived data
contain details of the object of the highest fidelity. This can in principle be used to extract and analyse
any desired physical properties on the macroscopic scale.

However, this method has some disadvantages. One often models macroscopic objects by means of their
constituent particles as large-scale interacting particle systems, where the known physical properties on the
microscopic scale are given by intermolecular forces. Thus, even for small macroscopic objects, the number of
particles that need to be modelled in order to obtain an accurate representation of the object is typically
very large. This inhibits the analytical evaluation of these models dramatically, except for a few
cases. Consequently, such models are usually analysed using computer simulations. Here, additionally, the
maximal step-size in numerical integration schemes that still ensures numerical stability is usually very
small. Both factors combined pose a huge obstacle in the scalability of simulations of interacting particle
systems in time and in space.

The goal is thus to develop mathematical methods that reduce the complexity of a given large scale interacting
particle system on the microscopic scale in a rigorous way so that the desired macroscopic properties of the
system are preserved. This process would free up resources that can be used to further scale the simulation in time as well as in space.

In this article, we start by summarising some of the mathematical methods that are used to reduce the
complexity of deterministic systems on the microscopic scale and will show how they preserve important
information on the macroscopic scale. These mathematical methods usually include some averaging procedures (coarse-graining techniques) or dimension reduction techniques. In the latter case, thermodynamic considerations often play an important role in the analysis of these systems. By construction, these mathematical methods only give macroscopic insight into the dynamics of the system. Therefore, in the main part of this article, we build on some of these mathematical methods and analyse the macroscopic properties of a simple fast-slow mechanical system in more detail by expanding the dynamics of the system to second-order and interpreting it from a thermodynamic point of view.

\subsection{Review of some averaging and dimension reduction techniques}

We focus on models which describe a macroscopic system by deterministic dynamics on the microscopic scale. This microscopic dynamics can be modelled in time and/or in space. Typical for these models
is the presence of a small scale-parameter \(\eps\). It represents the ratio of the fast dynamics on the microscopic scale and the slow dynamics on the macroscopic scale. A recurrent approach to study these
fast-slow systems is to consider the limit \(\eps\to0\) and thus derive an average or reduced model which is, in some sense, oblivious to the small-scale motion in the system. These procedures can broadly be classified into three different categories: non-projective, projective, and phenomenologically derived continuum mechanical methods.

While phenomenologically derived continuum mechanical methods aim to describe a macroscopic system as a continuous object without resorting to the microscopic scale, non-projective and projective methods aim to describe a macroscopic system by its most prominent properties that materialize in terms of some descriptive variables by studying the system on the microscopic scale. In general, projective methods start from a large-scale microscopic system and apply mathematical transformations to extract information about a handful of variables chosen \emph{a priori} such as volume, pressure, or density. Then the evolution of the microscopic system is largely confined to a low-dimensional submanifold, on which it can be described by these macroscopic variables. Non-projective methods often aim to describe the dynamics of the system by systematically deriving the most dominant motion of the whole system. Here, the dynamics of the whole system is reduced to the dynamics of a less complex system. In contrast to projective methods, the resulting variables are not necessarily given but appear in the process, e.g., through averaging, and can sometimes be given a physical interpretation \emph{a posteriori}, as shown in this article for a model problem.

Under the non-projective methods, the \emph{WKB method} is a classic and well known representative. As
described, for instance, in~\cite{Bender1978,Maslov1981a}, it is often used in the analysis of quantum
mechanical systems. It can be used to calculate an approximate solution to the stationary
Schrödinger equation, which is a second-order differential equation of the form
\begin{equation}
    \label{eq:a65}
    \eps^2\phi''(x) = Q(x)\phi(x),
\end{equation}
complemented with some initial conditions, where \(\eps\) is a small scale-parameter and \(Q\colon \Omega\to \RR\) is related to
the system's potential. The WKB method consists of an explicit ansatz for the oscillatory part of the
solution, which is necessary to analyse the dynamics for \(\eps> 0\). Equation~\eqref{eq:a65} can
approximately be solved with the WKB ansatz
\begin{equation*}
    \phi(x) \sim \exp\left[ \frac{1}{\eps}\sum_{n=0}^\infty \eps^n S_n(x) \right] \quad \text{for} \quad \eps\to0.
\end{equation*}
By comparing powers of \(\eps\) one derives a sequence of equations which determine \(S_0, S_1, \ldots\) and thus a representation of the solution at different scales.

The \emph{averaging method} is an alternative representative of the class of non-projective methods. It can be
used to derive the slow dynamics of solutions to ordinary differential equations if the evolution of the
system's DOF can be decomposed into the form
\begin{equation}
    \label{eq:a60}
    \dot{y} = f(y,z), \qquad \dot{z} = \eps^{-1} g(y,z),
\end{equation}
with \(y(0)=y_\ast\) and \(z(0)=z_\ast\), where \(y \colon \RR \to \RR^n\) describes the slow and \(z \colon \RR \to \RR^m\) the fast degrees of
freedom. In general, one is only interested in the dynamics of \(y\), while \(z\) is introduced to accurately model the dynamics on the microscopic scale. Systems of this kind can be used to model, for example, the evolution of molecules in the united atom representation~\cite{Schuette1997, Bornemann1997a, Bornemann1998a} where the parameter \(\eps\) represents the scale ratio between the fast molecular vibrations and the slow conformal motion of the molecule, or the development of long- and short-term weather phenomena~\cite{Imkeller2001, IMKELLER2002}, where \(\eps\) represents the scale ratio between the fast change of local weather phenomena and the slow development of the global climate.

A natural choice to reduce the complexity of the model~\eqref{eq:a60} is to average out the fast dynamics in the system. If the function \(f\) is periodic in \(z\) with period \(T\), then under very mild
assumptions, one can consider the averaged system
\begin{equation}
    \label{eq:a61}
    \dot{\bar{y}} = \bar{f}(\bar{y}), \qquad \text{where} \qquad  \bar{f}(\cdot)
    = \frac{1}{T} \int_0^{T} f(\cdot, s) \ud s,
\end{equation}
with \(\bar{y}(0)=y_\ast\). It can be shown (see for example~\cite{Arnold2006b, Sanders2007, Pavliotis2008})
that \(\bar{y}\) remains close to \(y\) for time-scales of order \(\mathcal{O}(1)\). Thus, system~\eqref{eq:a61} provides for a sufficiently small time-interval an approximate but less complex description of the dynamics of \(y\).

If the fast DOF are deterministic but sufficiently ``chaotic'', the scaled difference \(\eps^{-1/2} (y-\bar{y})\) can be interpreted in the limit \(\eps\to 0\) as Gaussian white noise, which can be analysed using probabilistic tools such as the central limit theorem or large deviation principles~\cite{Imkeller2001, IMKELLER2002}.

Another closely related non-projective methodology is the \emph{homogenisation of differential equations}~\cite{Murat1997, Cioranescu1999, Jikov1994}. Other than the averaging method, which is a coarse-graining method in time, the homogenisation method is a coarse-graining method in space. The idea is to simplify, for example, an elliptic partial
differential equation of the type
\begin{equation}
    \label{eq:a62}
    -\nabla \cdot (A(x/\eps)\nabla u_\eps(x)) = f(x)  \quad \text{for} \quad x\in \Omega,
    \qquad u_\eps =0 \quad \text{for} \quad x\in \partial\Omega,
\end{equation}
with \(f\in L^2(\Omega)\), \(\Omega\subset \RR^n\), where \(y\mapsto A(y)\in \RR^{n\times n}\) is \(1\)-periodic. By applying a two-scale
ansatz of the form
\begin{equation}
    \label{eq:a69}
    u_\eps(x)= u_0(x, \eps^{-1}x) +\eps u_1(x, \eps^{-1}x)+ \mathcal{O}(\eps^2),
\end{equation}
it can be shown (see, e.g.,~\cite{Pavliotis2008}) that \(u_0\) solves the homogenised partial differential equation
\begin{equation}
    \label{eq:a63}
    -\nabla \cdot (\bar{A}\nabla u_0(x)) = f(x)  \quad \text{for} \quad x\in \Omega,
    \qquad u_0 =0 \quad \text{for} \quad x\in \partial\Omega,
\end{equation}
for a computable homogenised conductivity tensor \(\bar{A}\). Without the \(\eps\)-dependent conductivity tensor \(A(x/\eps)\), system~\eqref{eq:a63} is less complex and thus its solution \(u_0\) can be easier derived than \(u_\eps\), the solution to the original system~\eqref{eq:a62}.

Other non-projective methods comprise the \emph{perturbation theory of integrable Hamiltonian
    systems}~\cite{Arnold2006b} or \emph{multiple-scale-asymptotics}~\cite{Bensoussan2011a}; modern
presentations for several of these approaches include~\cite{Pavliotis2008} and~\cite{Kuehn2015a}.

The \emph{Mori-Zwanzig framework}~\cite{Zwanzig2001,Mori1965} represents an example for a projective method. It has been extensively studied by Chorin \emph{et al.} under the name of ``optimal prediction'', for example in~\cite{Chorin1998, Chorin2001, Kupferman2004}. Other relevant references can be found in~\cite{Pavliotis2014}. The Mori-Zwanzig framework provides a way to study the reduced system for \(y\) by rewriting the deterministic system~\eqref{eq:a60} into a form which resembles a general Langevin
equation, where the dynamics of \(z\) is transformed into the stochastic component. In general, the Langevin equation for
\(y\) in the Mori-Zwanzig framework takes the form
\begin{equation}
    \label{eq:a64}
    \frac{d y}{d t}=h(y(t))+\int_{0}^{t} K(y(t-s), s) \ud s+\dot{W}_{t}.
\end{equation}
Here, the first term on the right-hand side is Markovian, the second term describes the possible memory effect of the process, and \(W_t\) denotes the stochastic process.

The Mori-Zwanzig framework can be seen as a generalisation of the averaging method. In particular, if the Mori-Zwanzig framework is applied to the deterministic system~\eqref{eq:a60}, then the Markovian term \(h\) and the memory kernel \(K\) in~\eqref{eq:a64} are \(\eps\)-dependent, i.e., \(h=h_\eps\) and \(K=K_\eps\). It is formally described in~\cite{Gottwald2016} that in this case, equation~\eqref{eq:a64} converges for \(\eps\to 0\) to the averaged equation~\eqref{eq:a61} derived with the averaging method.

In general, the stochastic process in equation~\eqref{eq:a64} describes a diffusion of the process and has the physical interpretation of thermal fluctuations. Thus, the dynamics of \(y\) can be interpreted in a thermodynamic sense.

Another framework in the class of projective methods, which is used to study non-equilibrium thermodynamic
processes modelled by fast-slow mechanical systems, is known as \emph{GENERIC} (General Equation for the Non-Equilibrium
Reversible-Irreversible Coupling,~\cite{Ottinger2005a, Grmela1997, Oettinger1997}). The idea is to find a projection operator so that the
system's DOF can be divided into a set of fast and slow variables. The fast variables are
collectively interpreted as the thermodynamic component of the system, while the slow variables are
interpreted as the mechanical component. The GENERIC is of the form
\begin{equation*}
    \frac{d x}{d t}=L \frac{\delta E}{\delta x}+M \frac{\delta S}{\delta x},
\end{equation*}
where \(x\) is the macroscopic quantity of interest, which changes in time driven by the non-equilibrium
thermodynamic processes, and \(E\) and \(S\) are potentials which have the physical meaning of energy and
entropy; \(L\) is a symplectic operator and \(M\) is positive semidefinite. If a system can be written in GENERIC
form, thermodynamic consistency is automatically ensured.

Finally, a macroscopic system can directly be described by \emph{continuum mechanics}~\cite{Tadmor2011}. Instead of modelling the system on the microscopic scale by interacting particles, in continuum mechanics the macroscopic object is
considered as a continuous object. Pivotal in this description is that the fundamental equations in continuum
mechanics are based on phenomenologically derived conservation laws. Their rigorous derivation from
interacting particle models is for many models an open problem.

\subsection{Context of this work}

While the thermodynamic characteristics of projective methods and the thermodynamic foundation of continuum
mechanics are well established theories that allow to analyse complex phenomena on the macroscopic scale, it is surprising that relatively little is known about non-projective upscaling methods and
their relation to thermodynamics~\cite{Grava2020,Chatterjee2020}. Insights into this relation are particularly important for molecular dynamic
simulations, since the aforementioned scaling problem is particularly pronounced in this field and one would
expect thermodynamic relations to hold, which could greatly speed up computations if suitably incorporated in
place of many-particle simulations of a solvent, for example.

To analyse the thermodynamic relation of non-projective upscaling methods, we study in this article a long-standing problem in the theory of mechanical systems, i.e., the ``strong confinement problem''~\cite{Rubin1957a, Arnold1989a, Takens1980,Froese2001}. In particular, we study a  simplified version of a model which was analysed in detail by Bornemann in~\cite{Bornemann1998a}. The original model was used to analyse the macroscopic dynamics of the four~\ce{CH}-groups of the butane molecule by deriving a homogenised model which is confined to a slow submanifold in the configuration space of the four~\ce{CH}-groups, using averaging methods in the form of homogenisation procedures.

The simplified version studied in this article
consists of a system of one fast and one slow particle, whose dynamics is governed by the
Lagrangian~\eqref{eq:03} below. This simplified model has the advantage that the notation can be kept to a
minimum while the most important results can still be conveyed to the reader. The results presented in this
article do generalise to a system of multiple fast and slow particles~\cite{Klar2021, Klar2021a}. The
low-dimensional presentation we have chosen here, however, makes the arguments more transparent, and the
extension to more complex situations is relatively straightforward. We use weak convergence methods in our
proofs similar to~\cite{Bornemann1998a}, see also~\cite{Evans2016} for a related approach. Moreover, because of
the simplicity of the model, it is possible to frame the analytical results in this article in a form using
two-scale convergence (cf.\ equation~\eqref{eq:a69}).

The interaction of one fast and one slow particle is one of the simplest models falling into the class of
mechanical systems studied by Bornemann in~\cite{Bornemann1998a}. It is also one of the simplest systems which
can potentially exhibit thermodynamic effects. Indeed, one of the core assumptions of thermodynamics is that
the system under consideration has a clear separation of scales (such as conformal motion described by
elasticity combined with fast oscillations described by temperature). The number of particles does not have to
be large or infinite. Notably, the physicist Paul Hertz developed a thermodynamic theory~\cite{Hertz1910a} for
Hamiltonian systems under a slow external perturbation. Specifically, he introduced an entropy, using a notion
of temperature as developed by Boltzmann. The book by Berdichevsky~\cite{Berdichevsky1997a} gives an excellent
introduction to this theory.

If one applies the theory of Hertz to system~\eqref{eq:03} below, one can interpret the dynamics associated
with the fast DOF as a fast subsystem that is slowly perturbed by the motion resulting from the slow
subsystem associated with the slow DOF. Hertz' theory then allows to describe the fast subsystem from a
thermodynamic point of view, using a notion of temperature \(T_\eps\), entropy \(S_\eps\), and external force
\(F_\eps\) (the force exerted by \(y_\eps\) on \(z_\eps\)). We reiterate that analogous findings hold for
many-particle interactions, as described below.

With \(\eps\) as a scale-parameter, we can analyse the system on different scales in time and space. In the
first part of this article, we rigorously derive a higher order asymptotic expansion of the slow dynamics of
the system using weak convergence techniques similar to~\cite{Bornemann1998a,Evans2016}. While the dynamics to leading-order is slow as already shown in~\cite{Bornemann1998a,Li2019b}, it
turns out that the dynamics to second-order can be decomposed into a slow component, describing the average
motion, and a fast component, describing fast oscillatory motion. The results
from the first part allow to similarly derive the second-order asymptotic expansion of the temperature, the
entropy, and the external force of the fast subsystem. It turns out that the leading-order as well as the
average dynamics to second-order satisfy equations which resemble the first and second law of
thermodynamics. This finding can potentially accelerate the simulation of slow dynamics in molecular dynamics
simulations to higher accuracy. Some numerical experiments can be found in~\cite{Klar2021, Klar2021a}.

\section{The model problem}

For a small scale-parameter \(0<\eps < \eps_0\), we study a family of mechanical systems described by the
Lagrangian
\begin{equation}
    \label{eq:03}
    \mathscr{L}_\eps(y_\eps, z_\eps,\dot{y}_\eps, \dot{z}_\eps)= \tfrac{1}{2}\dot{y}_\eps^2 +\tfrac{1}{2}\dot{z}_\eps^2
    - \tfrac{1}{2}\eps^{-2} \omega^2(y_\eps) z_\eps^2,
\end{equation}
on the two-dimensional Euclidean configuration space \(M=\RR^2\). This system is a simplified version of the
model problem introduced in~\cite[§1.2.1]{Bornemann1998a}. The corresponding Newtonian equations of motion
take the form
\begin{IEEEeqnarray}{rCl}
    \eqlabel{eq:04}
    \IEEEyesnumber \IEEEyessubnumber*
    \ddot{y}_\eps &=& -\eps^{-2}\omega(y_\eps)\omega'(y_\eps)z_\eps^2,\\
    \label{eq:05}
    \ddot{z}_\eps &=&-\eps^{-2} \omega^2(y_\eps) z_\eps.
\end{IEEEeqnarray}
We assume that \(\omega\in C^\infty(\mathbb{R})\) is a uniformly positive function, i.e., there is a constant
\(\omega_\ast>0\) such that
\begin{equation*}
    \label{eq:06}
    \omega(y) \geq \omega_\ast, \qquad \text{for all } y\in \mathbb{R}.
\end{equation*}
The \(\eps\)-independent initial values are
\begin{equation}
    \label{eq:08}
    y_\eps(0)=y_\ast,\qquad \dot{y}_\eps(0)=p_\ast,\qquad z_\eps(0)=0,\qquad \dot{z}_\eps(0)=u_\ast.
\end{equation}
Notice that \(y_\ast\in \RR\) can be chosen arbitrarily but the particular choice \(z_\eps(0)=0\) is necessary to ensure that the constant energy \(E_\eps\) of the system is independent of \(\eps\) and thus remains finite for all \(\eps\),
\begin{equation}
    \label{eq:09}
    E_\eps = \tfrac{1}{2}\dot{y}_\eps^2 + \tfrac{1}{2}\dot{z}_\eps^2 + \tfrac{1}{2}\eps^{-2}\omega^2(y_\eps) z_\eps^2
    = \tfrac{1}{2}p_\ast^2 + \tfrac{1}{2} u_\ast^2 = E_\ast.
\end{equation}

We are primarily interested in the time evolution of the slow DOF \(y_\eps\). Theorem 1
in~\cite[Chapter~I §2]{Bornemann1998a} shows that \(y_\eps\) converges in the limit \(\eps \to 0\) to a
function \(y_0\) in \(C^1([0,T])\), which is given as the solution to the second-order differential equation \(\ddot{y}_0=-\theta_\ast \omega'(y_0)\) with initial values \(y_0(0)=y_\ast, \dot{y}_0(0)=p_\ast\). The
constant \(\theta_\ast\) in the effective potential is of the form \(\theta_\ast=u_\ast^2/2\omega(y_\ast)\),
with \(y_\ast\) and \(u_\ast\) as in~\eqref{eq:08}. We will later see that \(\theta_\ast\) is proportional to
the action of the fast subsystem in the limit \(\eps\to0\). As a consequence, some information of the fast
subsystem is retained in the slow evolution of \(y_0\).

We extend the theory developed in~\cite{Bornemann1998a} by deriving rigorously the second-order asymptotic
expansion for the solution of the equations of motion~\eqref{eq:04} and interpret the corresponding expansion
of the energy~\eqref{eq:09} from a thermodynamic point of view. A crucial step in the derivation of these
expansions is the introduction of action-angle variables for the rapidly oscillating DOF
\((z_\eps,\dot{z}_\eps)\mapsto (\theta_\eps, \phi_\eps)\), which also involves a transformation of the
generalised momentum \(\dot{y}_\eps\mapsto p_\eps\) to preserve the symplectic structure on the phase-space as a whole.

\subsection{Main results}
\label{sec:main-res}
The main results in this work can be stated as follows:
\begin{enumerate}
    \item \label{thm:p1} There is a second-order asymptotic expansion of the variables
          \(y_\eps,\, p_\eps,\,\theta_\eps,\,\phi_\eps \) introduced in the previous paragraph and defined precisely
          in Section~\ref{sec:5}; this expansion is of the form
          \begin{IEEEeqnarray*}{rClClClCl}
              y_\eps &=& y_0 &+& \eps[\bar{y}_1]^\eps&+&\eps^2[\bar{y}_2]^\eps &+& \eps^2y_3^\eps,\\
              p_\eps &=& p_0 &+& \eps[\bar{p}_1]^\eps&+&\eps^2[\bar{p}_2]^\eps &+& \eps^2p_3^\eps,\\
              \theta_\eps &=& \theta_\ast &+& \eps[\bar{\theta}_1]^\eps&+&\eps^2[\bar{\theta}_2]^\eps &+& \eps^2\theta_3^\eps,\\
              \phi_\eps &=& \phi_0 &+& \eps[\bar{\phi}_1]^\eps&+&\eps^2[\bar{\phi}_2]^\eps &+& \eps^2\phi_3^\eps,
          \end{IEEEeqnarray*}
          where for \(i\in \{1,2\}\),
          \begin{IEEEeqnarray*}{rClClClCl+rClCl}
              [\bar{y}_i]^\eps &\coloneqq& \bar{y}_i &+& [y_i]^\eps&\weak{\ast}& \bar{y}_i \quad &\text{in}
              &\quad L^\infty([0,T]),&y_3^\eps &\to& 0 \quad &\text{in}&\quad C([0,T]),\\
              \phantom{}[\bar{p}_i]^\eps &\coloneqq& \bar{p}_i &+& [p_i]^\eps&\weak{\ast}& \bar{p}_i \quad &\text{in}
              &\quad L^\infty([0,T]),&p_3^\eps &\to& 0 \quad &\text{in}&\quad C([0,T]),\\
              \phantom{}[\bar{\theta}_i]^\eps &\coloneqq& \bar{\theta}_i &+& [\theta_i]^\eps&\weak{\ast}& \bar{\theta}_i \quad
              &\text{in}&\quad L^\infty([0,T]),&\theta_3^\eps &\to& 0 \quad &\text{in}&\quad C([0,T]),\\
              \phantom{}[\bar{\phi}_i]^\eps &\coloneqq& \bar{\phi}_i &+& [\phi_i]^\eps&\weak{\ast}& \bar{\phi}_i \quad
              &\text{in}&\quad L^\infty([0,T]),&\phi_3^\eps &\to& 0 \quad &\text{in}&\quad C([0,T]).
          \end{IEEEeqnarray*}
          In other words, for each variable the second-order asymptotic expansion is characterised --- to
          leading-order by Theorem 1 in~\cite{Bornemann1998a} --- to \(i\)-th order by a decomposition into a
          slow term, indicated by an overbar, which constitutes the average motion of the \(i\)-th order
          expansion, and a fast term, indicated by square brackets, which oscillate rapidly and converge
          weakly\(^\ast\) to zero in \(L^\infty([0,T])\) --- and by a residual term, indicated with a subscript three, which
          converges uniformly to zero in \(C([0,T])\). In particular, we show that
          \begin{equation*}
              [\bar{y}_1]^\eps = 0,\qquad [\bar{p}_1]^\eps = 0,\qquad  [\bar{\theta}_1]^\eps
              = [\theta_1]^\eps,\qquad [\bar{\phi}_1]^\eps = 0,
          \end{equation*}
          and that \((\bar{\phi}_2, \bar{\theta}_2, \bar{y}_2, \bar{p}_2)\) is given as the solution to the initial value problem~\eqref{eq:38},~\eqref{eq:a42} (Theorem~\ref{thm:2}). Moreover, the rapidly oscillating functions \([\theta_1]^\eps\), \([y_2]^\eps\), \([p_2]^\eps\), \([\theta_2]^\eps\), and
          \([\phi_2]^\eps\) are explicitly given in Definition~\ref{def:1}.

          Finally, we show that this expansion can be interpreted as a nonlinear version of a two-scale
          expansion, which we briefly introduce in Section~\ref{sec:brief-summary-two}.

    \item Using the framework of Hertz~\cite{Hertz1910a}, we define a temperature \(T_\eps\), an entropy
          \(S_\eps\), and an external force \(F_\eps\) for the fast subsystem (see Section~\ref{sec:6}). In
          combination with the analytic result discussed under item~\ref{thm:p1}, we decompose the total
          energy \(E_\eps\) into the energy associated with the fast subsystem \(E_\eps^\perp\), i.e.
          \begin{equation*}
              E_\eps^\perp = \tfrac{1}{2}\dot{z}_\eps^2 + \tfrac{1}{2}\eps^{-2}\omega^2(y_\eps) z_\eps^2,
          \end{equation*}
          and the residual energy \(E_\eps^\parallel= E_\eps - E_\eps^\perp\). We expand, similar to above,
          \(E_\eps^\perp\), \(E_\eps^\parallel\), \(T_\eps\), \(S_\eps\), and \(F_\eps\) into the form
          \begin{IEEEeqnarray*}{rClClClCl}
              E_\eps^\perp &=& E_0^\perp &+& \eps[\bar{E}_1^\perp]^\eps &+&\eps^2[\bar{E}_2^\perp]^\eps&+& \eps^2E_3^{\perp\eps},\\
              E_\eps^\parallel &=& E_0^\parallel &+& \eps[\bar{E}_1^\parallel]^\eps &+&\eps^2[\bar{E}_2^\parallel]^\eps&+& \eps^2E_3^{\parallel\eps},\\
              S_\eps &=& S_0 &+&\eps[\bar{S}_1]^\eps &+& \eps^2[\bar{S}_2]^\eps&+& \eps^2S_3^\eps,\\
              T_\eps &=& T_0 &+&  T_1^\eps,\\
              F_\eps &=& F_0 &+&  F_1^\eps,
          \end{IEEEeqnarray*}
          where \(T_1^\eps,F_1^\eps\to 0\) in \(C([0,T])\) and for \(i\in \{1,2\}\)
          \begin{IEEEeqnarray*}{rClClClCl+rClCl}
              [\bar{E}_i^\perp]^\eps &\coloneqq& \bar{E}_i^\perp &+& [E_i^\perp]^\eps&\weak{\ast}& \bar{E}_i^\perp \quad
              &\text{in}&\quad L^\infty([0,T]),&E_3^{\perp\eps} &\to& 0 \quad &\text{in}&\quad C([0,T]),\\
              \phantom{}[\bar{E}_i^\parallel]^\eps &\coloneqq& \bar{E}_i^\parallel &+& [E_i^\parallel]^\eps&\weak{\ast}
              & \bar{E}_i^\parallel \quad &\text{in}&\quad L^\infty([0,T]),&E_3^{\parallel\eps} &\to& 0 \quad &\text{in}&\quad C([0,T]),\\
              \phantom{}[\bar{S}_i]^\eps &\coloneqq& \bar{S}_i &+& [S_i]^\eps&\weak{\ast}& \bar{S}_i \quad
              &\text{in}&\quad L^\infty([0,T]),&S_3^\eps &\to& 0 \quad &\text{in}&\quad C([0,T]).
          \end{IEEEeqnarray*}
          The characterisation of the \(i\)-th order expansion is similar to above; moreover, it follows
          from~\eqref{eq:09} that
          \begin{equation*}
              E_\eps = E_0^\perp + E_0^\parallel = E_\ast ,\qquad [\bar{E}_1^\perp]^\eps + [\bar{E}_1^\parallel]^\eps = 0,
          \end{equation*}
          \begin{equation*}
              [\bar{E}_2^\perp]^\eps + [\bar{E}_2^\parallel]^\eps = 0, \qquad E_3^{\perp\eps} + E_3^{\parallel\eps} = 0.
          \end{equation*}
          In Section~\ref{sec:6} we show that
          \begin{equation*}
              [\bar{E}_1^\perp]^\eps = [E_1^\perp]^\eps,\qquad [\bar{E}_1^\parallel]^\eps = [E_1^\parallel]^\eps, \qquad [\bar{S}_1]^\eps = [S_1]^\eps,
          \end{equation*}
          and interpret the asymptotic expansion from a thermodynamic point of view. In particular, we show
          that to leading-order the entropy expression remains \emph{constant}, i.e., \(d S_0 = 0\), and
          consequently, the dynamics can be interpreted as an adiabatic thermodynamic process characterised by
          an energy relation that defines processes in thermodynamic equilibrium,
          \begin{equation*}
              d E_0^\perp = F_0 d y_0 + T_0 d S_0.
          \end{equation*}
          In contrast, we show that the averaged second-order dynamics, i.e., the dynamics in the
          weak\(^\ast\) limit, indicated by an overbar, represents a non-adiabatic thermodynamic process with
          an averaged \emph{nonconstant entropy}, \(d \bar{S}_2\neq 0\), that similar to above satisfies
          relations akin to equilibrium thermodynamics --- namely
          \begin{equation*}
              d \bar{E}_2^\perp = F_0 d \bar{y}_2 + T_0 d \bar{S}_2,
          \end{equation*}
          where \(\bar{S}_2\) indicates the averaged second-order entropy expression --- despite being beyond the limit \(\eps \to 0\). Finally, we show in Theorem~\ref{thm:3} that the evolution of \((\bar{y}_2, \bar{p}_2)\) is governed by equations which formally bear resemblance to Hamilton's canonical equations,
          \begin{IEEEeqnarray*}{rCl}
              \frac{d \bar{y}_2 }{dt} = \frac{\partial \bar{E}_2}{\partial p_0},&\qquad& \frac{\partial \bar{p}_2}{dt}
              = -\frac{\partial \bar{E}_2 }{\partial y_0},
          \end{IEEEeqnarray*}
          for \(\bar{E}_2 = \bar{E}_2^\perp + \bar{E}_2^\parallel\), which are complemented by the
          \(\eps\)-independent initial values
          \begin{equation*}
              \bar{y}_2(0)= -[y_2]^\eps(0), \qquad \bar{p}_2(0)= -[p_2]^\eps(0).
          \end{equation*}
\end{enumerate}

\section{The model problem in action-angle variables}

To study the dynamics of \(y_\eps\) and \(z_\eps\) for \(0<\eps<\eps_0\), a detailed asymptotic analysis is
required. An in-depth description for the case of multiple fast and slow DOF can be found
in~\cite{Klar2021}, which similarly uses ideas from~\cite{Bornemann1998a, Reich2000}. The proof sketches
given below are conceptually similar to those in~\cite{Klar2021}, but are more transparent due to less notational overhead. The idea is to transform the fast DOF into action-angle variables. To this
end, one first phrases the problem in Hamiltonian form. For this, one denotes by \((\eta_\eps, \zeta_\eps)\)
the canonical momenta corresponding to the positions \((y_\eps, z_\eps)\). Then the equations of
motion~\eqref{eq:04}, together with the velocity relations \(\dot{y}_\eps = \eta_\eps\) and
\(\dot{z}_\eps = \zeta_\eps\), are given by the canonical equations of motion belonging to the energy function
\begin{equation}
    \label{eq:106}
    E_\eps(y_\eps, \eta_\eps, z_\eps, \zeta_\eps) = \tfrac{1}{2}\eta_\eps^2 + \tfrac{1}{2}\zeta_\eps^2
    + \tfrac{1}{2}\eps^{-2}\omega^2(y_\eps)z_\eps^2.
\end{equation}

To take the oscillatory character of \(z_\eps\) into account, one introduces particular action-angle variables
\((\theta_\eps, \phi_\eps)\) for the fast DOF \((z_\eps, \zeta_\eps)\),
\begin{equation}
    \label{eq:105}
    z_\eps = \eps \sqrt{\frac{2\theta_\eps}{\omega(y_\eps)}}\sin(\eps^{-1}\phi_\eps),\qquad \zeta_\eps
    = \sqrt{2\theta_\eps \omega(y_\eps)}\cos(\eps^{-1}\phi_\eps),
\end{equation}
where we recall inequality~\eqref{eq:06}, i.e., \(\omega(y)\geq \omega_\ast >0\) for all \(y\in \RR\).

The transformation \((z_\eps, \zeta_\eps)\mapsto (\theta_\eps,\phi_\eps)\) can be found using the theory of generating functions~\cite[§48]{Arnold1989a}. Even though this transformation would be symplectic for fixed \(y_\eps\), this is not the case for the transformation of all phase-space variables \((y_\eps, \eta_\eps;z_\eps, \zeta_\eps)\mapsto (y_\eps, \eta_\eps; \theta_\eps,\phi_\eps)\). To ensure that the transformation of all phase-space variables remains symplectic, an additional transformation of the position \(y_\eps\) or the momentum \(\eta_\eps\) is required. If we
decide to keep the position variable unaffected by the transformation, the generating function takes the form
\begin{equation}
    \label{eq:a70}
    S_{\mathrm{gen}}(y_\eps, p_\eps, z_\eps, \phi_\eps) = p_\eps y_\eps+\tfrac{1}{2}\eps^{-1}\omega(y_\eps)z_\eps^2 \cot(\eps^{-1}\phi_\eps).
\end{equation}
The resulting transformation
\((y_\eps, \eta_\eps;z_\eps, \zeta_\eps)\mapsto (y_\eps, p_\eps; \phi_\eps, \theta_\eps)\) is symplectic on
the whole phase-space. Indeed, the energy function~\eqref{eq:106} transforms to the expression
\begin{equation}
    \label{eq:26}
    E_\eps = \frac{1}{2}p_\eps^2+ \theta_\eps \omega(y_\eps)+\eps \frac{\theta_\eps p_\eps
        \omega'(y_\eps)}{2\omega(y_\eps)}\sin(2\eps^{-1}\phi_\eps)+\frac{\eps^2}{8}
    \left( \frac{\theta_\eps \omega'(y_\eps)}{\omega(y_\eps)}
    \sin(2\eps^{-1}\phi_\eps) \right)^2,
\end{equation}
and the transformed DOF satisfy the equations of motion
\begin{equation*}
    \dot{\phi}_\eps = \frac{\partial E_\eps}{\partial \theta_\eps}, \qquad \dot{\theta}_\eps
    = -\frac{\partial E_\eps}{\partial \phi_\eps},\qquad \dot{y}_\eps = \frac{\partial E_\eps}{\partial p_\eps}, \qquad \dot{p}_\eps
    = - \frac{\partial E_\eps}{\partial y_\eps}.
\end{equation*}
After some calculations the equations of motion take the form
\begin{IEEEeqnarray*}{rCl}
    \eqlabel{eq:108}
    \IEEEyesnumber \IEEEyessubnumber*
    \label{eq:132}
    \dot{\phi}_\eps &=& \omega(y_\eps) +\eps \frac{p_\eps \omega'(y_\eps)}{2\omega(y_\eps)}  \sin(2\eps^{-1} \phi_\eps)
    +\eps^2 \frac{\theta_\eps \left( \omega'(y_\eps) \right)^2}{4 \omega^2(y_\eps)}\sin^2(2\eps^{-1}\phi_\eps),\IEEEeqnarraynumspace\\
    \label{eq:132a}
    \dot{\theta}_\eps &=& -\frac{ \theta_\eps p_\eps\omega'(y_\eps)}{\omega(y_\eps)}\cos(2\eps^{-1}\phi_\eps) -\eps
    \frac{\theta_\eps^2\left( \omega'(y_\eps) \right)^2}{4\omega^2(y_\eps)}\sin(4\eps^{-1}\phi_\eps),\\
    \label{eq:133}
    \dot{y}_\eps &=& p_\eps + \eps \frac{\theta_\eps \omega'(y_\eps)}{2 \omega(y_\eps)} \sin(2\eps^{-1}\phi_\eps),\\
    \label{eq:134}
    \dot{p}_\eps &=& -\theta_\eps \omega'(y_\eps) +\eps \frac{\theta_\eps p_\eps \left(\omega'(y_\eps) \right)^2}{2 \omega^2(y_\eps)}
    \sin(2\eps^{-1}\phi_\eps) \\
    && \nonumber -\> \eps \frac{\theta_\eps p_\eps\omega''(y_\eps)}{2\omega(y_\eps)} \sin(2\eps^{-1}\phi_\eps) + \eps^2 \frac{\theta_\eps^2 \left( \omega'(y_\eps) \right)^3}{4\omega^3(y_\eps)} \sin^2(2\eps^{-1}\phi_\eps) \\
    && \nonumber - \> \eps^2
    \frac{\theta_\eps^2 \omega'(y_\eps)\omega''(y_\eps) }{4\omega^2(y_\eps)}\sin^2(2\eps^{-1}\phi_\eps).
\end{IEEEeqnarray*}
The initial values, as given in~\eqref{eq:08}, transform to
\begin{equation}
    \label{eq:131}
    \phi_\eps(0)=0, \qquad \theta_\eps(0)=\theta_\ast = \frac{u_\ast^2}{2\omega(y_\ast)},\qquad y_\eps(0)= y_\ast, \qquad p_\eps(0)= p_\ast.
\end{equation}

At this point, it becomes clear, how the governing Newtonian equations of motion~\eqref{eq:04} are related to equation~\eqref{eq:a60}, which forms the basis for many fast-slow systems usually analysed using averaging methods. That is, by introducing the slow DOF \(\mathsf{y}_\eps\coloneqq (\theta_\eps, y_\eps, p_\eps)\) and the fast DOF \(\mathsf{z}_\eps \coloneqq \eps^{-1}\phi_\eps\), the system of differential equations~\eqref{eq:108} takes the form
\begin{IEEEeqnarray*}{rCl}
    \dot{\mathsf{y}}_\eps &=& f_0(\mathsf{y}_\eps,\mathsf{z}_\eps) + \eps f_1(\mathsf{y}_\eps,\mathsf{z}_\eps) + \eps^2 f_2(\mathsf{y}_\eps,\mathsf{z}_\eps), \\
    \dot{\mathsf{z}}_\eps &=& \eps^{-1} g_{-1}(\mathsf{y}_\eps,\mathsf{z}_\eps) + g_0(\mathsf{y}_\eps,\mathsf{z}_\eps) + \eps g_1(\mathsf{y}_\eps,\mathsf{z}_\eps),
\end{IEEEeqnarray*}
which, except for some higher-order terms, coincides with equation~\eqref{eq:a60}.

\section{Second-order asymptotic expansion}
\label{sec:5}

In this section, we rigorously derive the second-order asymptotic expansion for the solution of the initial value problem~\eqref{eq:108},~\eqref{eq:131}. Let us denote the right-hand side of~\eqref{eq:108} by \(\mathcal{F}_\eps \colon \mathbb{R}^4\to \mathbb{R}^4\). Because \(\mathcal{F}_\eps\) is locally Lipschitz continuous, by the standard existence and uniqueness theory for ordinary differential equations, there exists a \(0<T<\infty \) such that the initial value problem~\eqref{eq:108},~\eqref{eq:131} has a unique solution
\((\phi_\eps, \theta_\eps, y_\eps, p_\eps)\) in \(C^\infty([0,T],\mathbb{R}^4)\), for fixed \(0<\eps<\eps_0\).

\subsection{Leading-order expansion}

For \(0< \eps < \eps_0\), let \((\phi_\eps, \theta_\eps, y_\eps, p_\eps)\) in \(C^\infty([0,T],\mathbb{R}^4)\)
be the unique solution of the initial value problem~\eqref{eq:108},~\eqref{eq:131}. We analyse a sequence
\(\{\phi_\eps\}, \{\theta_\eps\}, \{y_\eps\}, \{p_\eps\}\) of this solution for \(\eps\to0\). The right-hand
side of~\eqref{eq:108} is oscillatory and has in particular highly oscillatory leading-order terms. As a
consequence, the sequences \(\{d\phi_\eps/dt\}\), \(\{\theta_\eps\}\), \(\{dy_\eps/dt\}\), \(\{dp_\eps/dt\}\)
are bounded in the space \(C^{0,1}([0,T])\) of uniformly Lipschitz continuous functions, while sequences of
higher-order derivatives (in particular \(\{d^2\theta_\eps/dt^2\}\), which will require special attention in
the later part of this work) become unbounded as \(\eps\to 0\). It follows from the extended Arzel\`a-Ascoli
Theorem~\cite[Principle~4, Chapter~I §1]{Bornemann1998a} that we can extract a subsequence, not relabelled,
and functions \(\theta_0 \in C^{0,1}([0,T])\) and \(\phi_0, y_0, p_0 \in C^{1,1}([0,T])\), such that
\begin{IEEEeqnarray}{rClCrCl}
    \eqlabel{eq:59}
    \IEEEyesnumber \IEEEyessubnumber*
    \label{eq:46A}
    \phi_\eps \to \phi_0& \quad \text{in} \quad& C^1([0,T]),& \qquad& \ddot{\phi}_\eps \weak{\ast}\ddot{\phi}_0& \quad \text{in}
    \quad& L^\infty([0,T]),\\
    \label{eq:46B}
    \theta_\eps \to \theta_0& \quad \text{in} \quad& C([0,T]),& \qquad& \dot{\theta}_\eps \weak{\ast}\dot{\theta}_0& \quad \text{in}
    \quad& L^\infty([0,T]),\\
    y_\eps \to y_0& \quad \text{in} \quad& C^1([0,T]),& \qquad&\ddot{y}_\eps \weak{\ast}\ddot{y}_0& \quad \text{in} \quad&
    L^\infty([0,T]),\\
    \label{eq:47A}
    p_\eps \to p_0& \quad \text{in} \quad& C^1([0,T]),& \qquad& \ddot{p}_\eps \weak{\ast}\ddot{p}_0& \quad \text{in} \quad& L^\infty([0,T]).
\end{IEEEeqnarray}
By taking the limit \(\eps \rightarrow 0\) in equation~\eqref{eq:132},~\eqref{eq:133}, and~\eqref{eq:134} we deduce that \(\dot{\phi}_0 = \omega(y_0)\), \(\dot{y}_0 = p_0\), and \(\dot{p}_0 = -\theta_0 \omega'(y_0)\). Moreover, from equation~\eqref{eq:132a} it can be read off that \(\dot{\theta}_\eps\) is rapidly oscillating around zero. By observing that the weak\(^\ast\) convergence in \(L^\infty([0,T])\) helps to ignore rapid fluctuations of functions, property~\eqref{eq:46B} can be used to deduct that \(\dot{\theta}_0=0\) and in particular, that \(\theta_0\equiv \theta_\ast\) (compare with~\eqref{eq:131}).

Finally, since the right-hand side of the limit equations
\begin{equation}
    \label{eq:4}
    \dot{\phi}_0 = \omega(y_0),\qquad \dot{\theta}_0 =0,\qquad \dot{y}_0 = p_0, \qquad \dot{p}_0 = -\theta_\ast \omega'(y_0),
\end{equation}
do not depend on a chosen subsequence, we can discard the extraction of subsequences altogether~\cite[Principle~5, Chapter~I §1]{Bornemann1998a}.

\subsection{Reformulation of the governing equations}
\label{sec:Reform-govern-equat}

For the following part of this work, it is convenient to introduce a notation which simplifies the system of
differential equations~\eqref{eq:108}; namely, for \(0\leq \eps < \eps_0\) and \(k,l \in \mathbb{N}_0\) we define the expression \(L_\eps \coloneqq \log\left( \omega(y_\eps) \right)\) and, based on this,
\begin{equation*}
    D^k_t D^l_y L_\eps \coloneqq \frac{d^k}{dt^k}\frac{d^l L_\eps}{d y_\eps^l}.
\end{equation*}

Then the system of differential equations~\eqref{eq:108} can be written as
\begin{IEEEeqnarray}{rCl}
    \eqlabel{eq:58}
    \IEEEyesnumber\IEEEyessubnumber*
    \label{eq:13}
    \dot{\phi}_\eps &=& \omega(y_\eps) + \frac{\eps}{2} D_t L_\eps \sin \left( 2\eps^{-1}\phi_\eps\right), \\
    \label{eq:14}
    \dot{\theta}_\eps&=& - \theta_\eps D_t L_\eps \cos(2\eps^{-1}\phi_\eps),\\
    \label{eq:15}
    \dot{y}_\eps &=& p_\eps + \frac{\eps }{2} \theta_\eps D_y L_\eps \sin(2\eps^{-1}\phi_\eps),\\
    \label{eq:16}
    \dot{p}_\eps &=& -\theta_\eps \omega'(y_\eps) - \frac{\eps}{2} \theta_\eps D_t D_y L_\eps \sin(2\eps^{-1}\phi_\eps).
\end{IEEEeqnarray}

\subsection{First- and second-order expansion}
To analyse the dynamics of the model problem away from the limit \(\eps\to0\), a higher-order asymptotic
expansion in \(\eps\) is required.

We first define particular functions that appear in the first- and second-order expansion before we state the
theorem that embodies the first main result (item~\ref{thm:p1}) of Section~\ref{sec:main-res}.

\begin{definition}
    \label{def:1}
    Let \((\phi_\eps, \theta_\eps, y_\eps, p_\eps)\) be the solution to the initial value problem~\eqref{eq:108},~\eqref{eq:131} and
    \((\phi_0,\theta_0,y_0,p_0)\) be the solution to the initial value problem~\eqref{eq:4},~\eqref{eq:131} such that~\eqref{eq:59} holds. With the notation introduced in Section~\ref{sec:Reform-govern-equat}, we define the functions
    \begin{equation}
        \label{eq:36b}
        \theta_1^\eps\coloneqq \frac{\theta_\eps-\theta_\ast}{\eps},
    \end{equation}
    and
    \begin{equation}
        \label{eq:36a}
        \phi_2^\eps \coloneqq
        \frac{\phi_\eps-\phi_0}{\eps^2},
        \qquad y_2^\eps \coloneqq \frac{y_\eps-y_0}{\eps^2},\qquad p_2^\eps \coloneqq \frac{p_\eps-p_0}{\eps^2},
        \qquad\theta_2^\eps \coloneqq
        \frac{\theta_1^\eps-[\theta_1]^\eps}{\eps},
    \end{equation}
    \begin{IEEEeqnarray*}{rCl+rCl}
        [\theta_1]^\eps &\coloneqq& -\frac{\theta_\ast D_t L_0}{2\omega(y_0)}\sin(2\eps^{-1}\phi_0),& [\phi_2]^\eps
        &\coloneqq& -\frac{D_t L_0}{4\omega(y_0)}\cos(2\eps^{-1}\phi_0),\\
        \phantom{}[y_2]^\eps&\coloneqq& -\frac{\theta_\ast D_y L_0}{4\omega(y_0)}\cos(2\eps^{-1}\phi_0),&  [p_2]^\eps
        &\coloneqq& \frac{d}{dt}\left( \frac{\theta_\ast D_y L_0}{4\omega(y_0)} \right)\cos(2\eps^{-1}\phi_0),
    \end{IEEEeqnarray*}
    and
    \begin{IEEEeqnarray*}{rCl}
        \phantom{}[\theta_2]^\eps &\coloneqq& -\theta_\ast D_y L_0 [y_2]^\eps - \frac{p_0}{\omega(y_0)} [p_2]^\eps
        + \frac{\theta_\ast^2 (D_y L_0)^2}{16 \omega(y_0)}\cos(4\eps^{-1}\phi_0) \\
        && - \> \frac{\theta_\ast D_t L_0 }{\omega(y_0)}\bar{\phi}_2\cos(2\eps^{-1}\phi_0).
    \end{IEEEeqnarray*}
\end{definition}

The functions \(\theta_1^\eps\), \(\phi_2^\eps\), \(y_2^\eps\), and \(p_2^\eps\) as defined
in~\eqref{eq:36b} and~\eqref{eq:36a} describe scaled versions of the residual motion of the originally given
DOF and their homogenised versions. The corresponding subscript indicates the scaling order and
marks their relevance in the first- and second-order expansion. To determine the relevant term in the
second-order expansion for \(\theta_\eps\), we similarly define by \(\theta_2^\eps\) the scaled residual
motion of \(\theta_\eps\) and its first-order expansion, which is derived in a two-step procedure via
\(\theta_1^\eps\). As indicated before in item~\ref{thm:p1} in Section~\ref{sec:main-res}, the second-order
expansions consist of oscillating and non-oscillating terms. The oscillating terms are denoted by expressions
in square brackets. They oscillate rapidly around zero and thus satisfy
\begin{equation}
    \label{eq:56}
    [\theta_1]^\eps, [\phi_2]^\eps, [y_2]^\eps, [p_2]^\eps, [\theta_2]^\eps \weak{\ast} 0 \quad \text{in} \quad L^\infty([0,T]).
\end{equation}
The non-oscillatory terms are characterised in the
following theorem, which is the main analytic result of this article. One key result is that the
non-oscillatory terms of the second-order expansion, marked by an overbar and subscript \(2\), satisfy a
system of ordinary differential equations.

\begin{theorem}
    \label{thm:2}
    Let \((\phi_\eps, \theta_\eps, y_\eps, p_\eps)\) be the solution to the initial value problem~\eqref{eq:108},~\eqref{eq:131} and
    \((\phi_0,\theta_0,y_0,p_0)\) be the solution to the initial value problem~\eqref{eq:4},~\eqref{eq:131} such that~\eqref{eq:59} holds. Then, the functions specified in Definition~\ref{def:1} satisfy
    \begin{IEEEeqnarray*}{rClCl+rClCl}
        \theta_1^\eps -[\theta_1]^\eps &\to& 0 &\quad \text{in} \quad& C([0,T]), & \frac{d}{dt}\left( \theta_1^\eps
        -[\theta_1]^\eps \right) &\weak{\ast}&0&\quad \text{in}\quad& L^\infty([0,T]),\IEEEeqnarraynumspace\\
        \phi_2^\eps -[\phi_2]^\eps &\to& \bar{\phi}_2 &\quad \text{in} \quad& C([0,T]),& \frac{d}{dt}\left( \phi_2^\eps
        -[\phi_2]^\eps \right)&\weak{\ast}&\frac{d \bar{\phi}_2}{dt}&\quad \text{in}\quad& L^\infty([0,T]),\\
        y_2^\eps -[y_2]^\eps &\to& \bar{y}_2 &\quad \text{in} \quad& C([0,T]),& \frac{d}{dt}
        \left( y_2^\eps -[y_2]^\eps \right)&\weak{\ast}&\frac{d \bar{y}_2}{dt}&\quad \text{in}\quad &L^\infty([0,T]),\\
        p_2^\eps -[p_2]^\eps &\to& \bar{p}_2 &\quad \text{in} \quad& C([0,T]),& \frac{d}{dt}
        \left( p_2^\eps -[p_2]^\eps \right)&\weak{\ast}&\frac{d \bar{p}_2}{dt}&\quad \text{in}\quad& L^\infty([0,T]),
    \end{IEEEeqnarray*}
    and
    \begin{equation*}
        \theta_2^\eps -[\theta_2]^\eps \to \bar{\theta}_2 \quad \text{in}\quad  C([0,T]),
    \end{equation*}
    where \((\bar{\phi}_2, \bar{\theta}_2, \bar{y}_2, \bar{p}_2)\) is the unique solution to the inhomogeneous
    linear system of differential equations
    \begin{IEEEeqnarray}{rCl}
        \eqlabel{eq:38}
        \IEEEyesnumber \IEEEyessubnumber*
        \label{eq:39A}
        \frac{d \bar{\phi}_2}{dt} &=&  \omega'(y_0)\bar{y}_2  + \frac{\theta_\ast (D_y L_0)^2}{8} - \frac{(D_t L_0)^2}{8
            \omega(y_0)},\\
        \label{eq:39}
        \frac{d \bar{\theta}_2}{dt}&=&\frac{d}{dt} \frac{\theta_\ast (D_t L_0)^2}{8 \omega^2(y_0)},\\
        \label{eq:40}
        \frac{d \bar{y}_2}{dt} &=& \bar{p}_2 - \frac{\theta_\ast D_y L_0 D_t L_0 }{4\omega(y_0)},\\
        \label{eq:41}
        \frac{d \bar{p}_2}{dt} &=& -\omega'(y_0) \bar{\theta}_2- \theta_\ast \omega''(y_0) \bar{y}_2-\frac{\theta_\ast^2
            D_y L_0 D_y^2 L_0}{8}+\frac{\theta_\ast  D_t L_0 D_t D_y L_0}{4\omega(y_0)},
    \end{IEEEeqnarray}
    with \(\eps\)-independent initial values
    \begin{IEEEeqnarray}{rCl}
        \eqlabel{eq:a42}
        \IEEEyesnumber \IEEEyessubnumber*
        \bar{\phi}_2(0)&=& -[\phi_2]^\eps(0),\qquad\bar{\theta}_2(0)=-[\theta_2]^\eps(0),\\
        \bar{y}_2(0)&=& -[y_2]^\eps(0), \qquad \bar{p}_2(0)= -[p_2]^\eps(0).
    \end{IEEEeqnarray}
\end{theorem}

The result in Theorem~\ref{thm:2} is central for this article in two different ways. Firstly, it will be crucial
for the thermodynamic interpretation in the second part of this article. Secondly, it is interesting for
computational purposes. That is, in simulating a natural evolution of a light particle coupled to a heavy
particle, their mass ratio \(\eps\) will be small but finite and enters into the underlying model through
potentials of different strengths. The result above says that rather than solving the coupled system directly,
which is restricted to a small step size to ensure numerical stability, the approximation to second-order can
be computed by combining explicitly known oscillatory functions (as given in Definition~\ref{def:1}) with the
solution of an inhomogeneous linear system of differential equations, as given in Theorem~\ref{thm:2}.

\subsubsection{Proof of Theorem~\ref{thm:2}}
\label{sec:2}

The detailed proof of Theorem~\ref{thm:2} was already carried out, in the case of multiple fast and slow
DOF in~\cite{Klar2021}. For this reason, we will only summarise the essential steps and
indicate the differences. The simpler notation in the present article makes the presentation more transparent.

It is first shown, that the sequences of functions \(\{\theta_1^\eps\}, \{\phi_2^\eps\}, \{y_2^\eps\}\), and
\(\{p_2^\eps\}\) are uniformly bounded in \(L^\infty([0,T])\). This follows directly from the system of
differential equations~\eqref{eq:58} and Gronwall's inequality. Moreover, one shows that the sequence of
functions \(\{\theta_2^\eps\}\) is also uniformly bounded in \(L^\infty([0,T])\). Other than
in~\cite{Klar2021}, which requires lengthy calculations, this follows directly from the second-order energy
(equation~\eqref{eq:a67} and~\eqref{eq:a68}) and the uniform boundedness of
\(\{\theta_1^\eps\}, \{\phi_2^\eps\}, \{y_2^\eps\}\), and \(\{p_2^\eps\}\).

Since we introduced the action-angle variables, it is possible to calculate the high-frequency terms
\([\theta_1]^\eps, [\phi_2]^\eps, [y_2]^\eps, [p_2]^\eps\), and \([\theta_2]^\eps\) through integration by
parts.

By taking the time derivative of the functions \( \phi_2^\eps -[\phi_2]^\eps\), \(y_2^\eps- [y_2]^\eps\) and
\(p_2^\eps-[p_2]^\eps\), we obtain
\begin{IEEEeqnarray*}{rCl}
    \frac{d}{dt}\left( \phi_2^\eps -[\phi_2]^\eps \right)&=&  \frac{\omega(y_\eps)-\omega(y_0)}{\eps^2}
    + \frac{d}{dt}\left( \frac{D_t L_\eps}{4\dot{\phi}_\eps} \right)\cos(2\eps^{-1}\phi_\eps) \\
    && - \frac{d}{dt}
    \left([\phi_2]^\eps +\frac{D_t L_\eps}{4\dot{\phi}_\eps}\cos(2\eps^{-1}\phi_\eps) \right),\\
    \frac{d}{dt}\left(y_2^\eps- [y_2]^\eps \right)&=&\frac{p_\eps - p_0}{\eps^2}+\frac{d}{dt}
    \left( \frac{\theta_\eps D_y L_\eps}{4\dot{\phi}_\eps} \right)\cos(2\eps^{-1}\phi_\eps) \\
    && - \frac{d}{dt}\left( [y_2]^\eps
    + \frac{\theta_\eps  D_y L_\eps}{4\dot{\phi}_\eps}\cos(2\eps^{-1}\phi_\eps) \right),\\
    \IEEEyesnumber\label{eq:52}
    \frac{d}{dt}\left(p_2^\eps-[p_2]^\eps \right) &=& -\theta_\ast \frac{\omega^\prime(y_\eps)-\omega^\prime(y_0)}{\eps^2}
    -\frac{\theta_1^\eps
    - [\theta_1]^\eps}{\eps} \omega^\prime(y_\eps)\\
    && -\> \frac{d}{dt}\left( \frac{\theta_\eps D_tD_y L_\eps}{4\dot{\phi}_\eps} \right)\cos(2\eps^{-1}\phi_\eps) \\
    && -\frac{d}{dt}\left( [p_2]^\eps_1 - \frac{\theta_\eps D_tD_yL_\eps}{4\dot{\phi}_\eps}\cos(2\eps^{-1}\phi_\eps) \right)\\
    &&+ \>\frac{d}{dt}\left( \frac{\theta_\ast D_t L_0}{4\omega^2(y_0)} \omega^\prime(y_\eps) \right)\cos(2\eps^{-1}\phi_0) \\
    && - \frac{d}{dt}\left( [p_2]^\eps_2+\frac{\theta_\ast D_t L_0}{4\omega^2(y_0)}\omega^\prime(y_\eps)\cos(2\eps^{-1}\phi_0)\right),
\end{IEEEeqnarray*}
where we write \([p_2]^\eps = [p_2]^\eps_1 + [p_2]^\eps_2\) with
\begin{equation*}
    [p_2]^\eps_1 \coloneqq \frac{\theta_\ast D_t D_y L_0}{4\omega(y_0)}\cos(2\eps^{-1}\phi_0),\qquad [p_2]^\eps_2 \coloneqq
    -\frac{\theta_\ast D_t L_0  D_y L_0}{4\omega(y_0) }\cos(2\eps^{-1}\phi_0).
\end{equation*}
For the derivation of equation~\eqref{eq:52}, we note that in
\begin{IEEEeqnarray*}{rCl}
    \frac{dp_2^\eps}{dt} &=& -\theta_\ast\frac{\omega^\prime(y_\eps)-\omega^\prime(y_0)}{\eps^2}
    - \frac{\theta_\eps-\theta_\ast}{\eps^2} \omega^\prime(y_\eps) - \frac{d}{dt}
    \left( \frac{\theta_\eps D_tD_y L_\eps}{4\dot{\phi}_\eps} \right)\cos(2\eps^{-1}\phi_\eps) \\
    && +\frac{d}{dt}
    \left( \frac{\theta_\eps D_tD_yL_\eps}{4\dot{\phi}_\eps}\cos(2\eps^{-1}\phi_\eps) \right),
\end{IEEEeqnarray*}
we can rewrite the second term on the right-hand side by introducing \([\theta_1]^\eps\), i.e.,
\begin{IEEEeqnarray*}{rCl}
    \frac{\theta_\eps-\theta_\ast}{\eps^2} \omega^\prime(y_\eps) &=& \frac{\theta_1^\eps
    - [\theta_1]^\eps}{\eps} \omega^\prime(y_\eps)- \frac{d}{dt}\left( \frac{\theta_\ast D_t L_0}{4\omega^2(y_0)}
    \omega^\prime(y_\eps) \right)
    \cos(2\eps^{-1}\phi_0) \\
    && +\frac{d}{dt}\left( \frac{\theta_\ast D_t L_0}{4\omega^2(y_0)}
    \omega^\prime(y_\eps)\cos(2\eps^{-1}\phi_0) \right).
\end{IEEEeqnarray*}

As described in~\cite{Klar2021}, we infer that the sequences
\(\{\phi_2^\eps -[\phi_2]^\eps\},\{y_2^\eps -[y_2]^\eps\}\), and \(\{p_2^\eps -[p_2]^\eps\}\) are bounded in
\(C^{0,1}([0,T])\). The claim follows after successive applications of the extended Arzel\`a-Ascoli
theorem~\cite[Principle~4, Chapter~I §1]{Bornemann1998a}. For the reader's convenience, we will exemplify the sketch of proof for \(y_\eps\). We recall that
\begin{IEEEeqnarray*}{rCl}
    \frac{d}{dt}\left(y_2^\eps- [y_2]^\eps \right)&=&\frac{p_\eps - p_0}{\eps^2}+\frac{d}{dt}
    \left( \frac{\theta_\eps D_y L_\eps}{4\dot{\phi}_\eps} \right)\cos(2\eps^{-1}\phi_\eps) \\
    && - \frac{d}{dt}\left( [y_2]^\eps
    + \frac{\theta_\eps  D_y L_\eps}{4\dot{\phi}_\eps}\cos(2\eps^{-1}\phi_\eps) \right),
\end{IEEEeqnarray*}
where we take the weak\(^\ast\) limit of the right-hand side. The first term on the right-hand side is uniformly bounded in
\(L^\infty([0,T])\). It follows that there exists a function \(\bar{p}_2\in L^\infty([0,T])\) and a
subsequence such that \(p_2^\eps \weak{\ast}\bar{p}_2\) in \(L^\infty([0,T])\). The second term on the
right-hand side converges weakly\(^\ast\) to the second term on the right-hand side of equation~\eqref{eq:40}, which follows from Lemma 5.7 in~\cite{Klar2021}. The weak\(^\ast\) limit of this term
is nonzero because the terms that appear from the amplitude after taking the time-derivative are in resonance
with the cosine term, thus leading to a nonzero contribution in the weak\(^\ast\) limit. Finally, the last
term on the right-hand side converges weakly\(^\ast\) to zero by construction.

The weak\(^\ast\) limits for \( \phi_2^\eps -[\phi_2]^\eps\) and \(p_2^\eps-[p_2]^\eps\) can be derived in a similar way. The form of \([\bar{\theta}_2]^\eps\) can then be derived through expansion of the energy term.

\subsection{Interpretation of the asymptotic expansion in the two-scale convergence framework}
\label{sec:brief-summary-two}

The convergence results in Theorem~\ref{thm:2} exhibit scale separations that are characteristic of the theory
of two-scale convergence. In this section, we give a summary of the theory and introduce a nonlinear version
of two-scale convergence, which can be used to reformulate the results derived in Theorem~\ref{thm:2}.

\subsubsection{Two-scale convergence}
The theory of two-scale convergence was first introduced by Nguetseng~\cite{Nguetseng1989a}. We follow here
the presentation in~\cite{Allaire1992b, Visintin2006b}, though restricted to the one-dimensional
case. We denote by \(\mathcal{S}\) the set \(S\coloneqq[0,1)\) equipped with the topology of the
\(1\)-dimensional torus, and identify any function on \(\mathcal{S}\) with its \(1\)-periodic extension on \(\mathbb{R}\).

In general, a bounded sequence \(\{u_\eps\}\) of functions in \(L^2(\Omega)\) is said to \emph{weakly
    two-scale converge} to \(u\in L^2(\Omega\times \mathcal{S})\), symbolically indicated by
\(u_\eps \xrightharpoonup[2]{} u\), if and only if
\begin{equation}
    \label{eq:42}
    \lim_{\eps\to 0}\int_\Omega u_\eps(t)\psi\left( t,\eps^{-1}t\right) \ud t
    = \iint_{\Omega\times S} u(t,s) \psi(t,s)\ud t\ud s,
\end{equation}
for any smooth function \(\psi\colon \mathbb{R}\times \mathbb{R}\to \mathbb{R}\) which is \(S\)-periodic with respect to the second argument. Typically, these \(u_\eps\) are of the form \(u_\eps(t)=v(t,\eps^{-1}t)\) for some function \(v\) of two arguments, which is periodic in the second argument.

\subsubsection{Two-scale decomposition}
For any \(\eps>0\), one can decompose a real number \(t\in \RR\) as
\(t = \eps [\mathcal{N}(t/\eps)+ \mathcal{R}(t/\eps)]\), where
\begin{equation*}
    \mathcal{N}(t)\coloneqq \max \{n\in \mathbb{Z}\colon n \leq t\},\quad \mathcal{R}(t)\coloneqq t
    - \mathcal{N}(t)\in \mathcal{S}.
\end{equation*}
Visually, \(\mathcal{N}\) is the floor function and \(\mathcal{R}\) is the sawtooth wave function. If \(\eps\) is the ratio between two disparate scales, \(\mathcal{N}(t/\eps)\) and \(\mathcal{R}(t/\eps)\) may
then be regarded as a coarse-scale and a fine-scale variable, respectively. Besides this two-scale
decomposition, one defines a \emph{two-scale composition function}:
\begin{equation*}
    h_\eps(t,s)\coloneqq \eps \mathcal{N}(t/\eps) + \eps s\qquad \forall(t,s) \in \mathbb{R}\times \mathcal{S},\qquad \forall\eps>0.
\end{equation*}

The two-scale composition function can be written as \(h_\eps(t,s)=t+\eps[s-\mathcal{R}(t/\eps)]\). Since, for all \(t\in \RR\), \(\mathcal{R}(t)\in \mathcal{S}\), one has \(\eps \mathcal{R}(t/\eps)\to 0\) uniformly in \(t\) and thus
\begin{equation*}
    h_\eps(t,s)\to t \quad \text{uniformly in } \mathbb{R}\times \mathcal{S},\text{ as }  \eps\to 0.
\end{equation*}

With the introduction of \(h_\eps\) one can define an equivalent definition of two-scale convergence. I.e.,
one has for a sequence \(\{u_\eps\}\) in \(L^2(\mathbb{R})\) that
\begin{equation*}
    u_\eps \xrightharpoonup[2]{} u \quad \textnormal{in}\quad L^2(\mathbb{R}\times \mathcal{S})\quad \Leftrightarrow \quad
    u_\eps\circ h_\eps \xrightharpoonup[]{} u \quad \textnormal{in}\quad L^2(\mathbb{R}\times \mathcal{S}).
\end{equation*}

For any domain \(\Omega\subset \mathbb{R}\), two-scale convergence in \(L^2(\Omega\times \mathcal{S})\) is
then defined by extending functions to \(\mathbb{R}\setminus \Omega\) with vanishing value.

For \(u_\eps \in L^2(\Omega)\) it is shown in~\cite{Visintin2006b}, that this definition of two-scale convergence is equivalent to the
definition in~\eqref{eq:42}. However, it is more versatile. In particular, it allows defining two-scale
convergence in \(C([0,T]\times \mathcal{S})\).

\subsubsection{Two-scale convergence in \texorpdfstring{\(\bm{C([0,T]\times \mathcal{S})}\)}{C0}}

Some modifications are needed to extend the definition of two-scale convergence to \(C([0,T]\times \mathcal{S})\), for in general the function \(u_\eps\circ h_\eps\) is discontinuous with
respect to \(t\in \mathbb{R}\) and \(s\in \mathcal{S}\), even if \(u_\eps\) is continuous. One therefore
replaces \(u_\eps\circ h_\eps\) by a continuous function,
\(\mathcal{L}_\eps u_\eps\coloneqq (J \circ I_\eps)(u_\eps\circ h_\eps)\), constructed via linear interpolation with
respect to each argument. The details of this linear interpolation can be found in~\cite{Visintin2006b}.

One then says that \(u_\eps\) \emph{strongly two-scale converges} to \(u\) in \(C([0,T]\times\mathcal{S})\),
symbolically indicated by \(u_\eps \xrightarrow[2]{} u\), if and only if
\begin{equation*}
    \mathcal{L}_\eps u_\eps \to u \quad \text{in} \quad C([0,T]\times
    \mathcal{S}).
\end{equation*}

\subsubsection{Nonlinear two-scale convergence}

The two-scale convergence theory is a generalisation of the weak convergence theory which retains
information about the oscillatory character of \(u_\eps\) in the limit \(\eps \to 0\). For instance, we have
\(\sin(2\pi\eps^{-1}t)\xrightharpoonup[2]{}\sin(2\pi s)\) in \(L^2([0,T]\times \mathcal{S})\). Notice however, that with test functions in form of the Fourier basis functions \(\psi_k(t, \eps^{-1}t)\coloneqq \exp(2\pi i k\eps^{-1}t)\) one has for all \(k\in \mathbb{Z}\)
\begin{equation*}
    \sin\left( 2\pi\eps^{-1}\varphi(t)\right)\psi_k(t,\eps^{-1}t) \weak{} 0 \quad \text{in} \quad L^2([0,T]\times \mathcal{S}),
\end{equation*}
for any nonlinear  \(C^\infty\)-Diffeomorphism \(\varphi \colon [0,T]\to [0, \varphi(T)]\), and thus \(\sin\left( 2\pi\eps^{-1}\varphi(t)\right) \xrightharpoonup[2]{} 0\) in \(L^2([0,T]\times \mathcal{S})\). To derive a non-zero two-scale limit in such a case, we introduce a nonlinear change of coordinates that
temporarily annihilates the nonlinearity so that the standard two-scale limit can be taken before it is
reintroduced into the two-scale limit.

\begin{definition}[Two-scale convergence with respect to $\varphi$]
    \label{def:2}
    Let \(\varphi\colon [0,T]\to [0, \varphi(T)]\) be a \(C^\infty\)-Diffeomorphism, \(\{u_\eps\}\subset C([0,T])\) and \(u\in C([0,T]\times \mathcal{S})\). We say that \(u_\eps\)
    \emph{two-scale converges} with respect to \(\varphi\) to \(u\) in
    \(C([0,T]\times \mathcal{S})\) if \(u_\eps\circ \varphi\) two-scale converges to \(u\circ(\varphi, \mathrm{Id})\) in \(C([0,\varphi(T)]\times \mathcal{S})\), i.e.,
    \begin{equation*}
        u_\eps \xrightarrow[2]{\varphi} u \quad \text{in} \quad C([0,T]\times \mathcal{S})
    \end{equation*}
    if and only if
    \begin{equation*}
        u_\eps( \varphi( r)) \xrightarrow[2]{} u(\varphi(r), s) \quad \text{in} \quad C([0,\varphi^{-1}(T)]\times \mathcal{S}).
    \end{equation*}
\end{definition}
With this definition at hand, we can express the uniform convergence results of Theorem~\ref{thm:2} using the
notation of two-scale convergence. Rephrasing the weak\(^\ast\) convergence results in Theorem~\ref{thm:2} in a similar way requires more notation.

\begin{proposition}
    \label{cor:1}
    With the notation introduced in Definition~\ref{def:2}, the uniform convergences in Theorem~\ref{thm:2}
    are equivalent to the following strong two-scale convergences with respect to \(\varphi(r)\coloneqq \phi_0^{-1}(\pi r)\) for \(r\in [0,\pi^{-1}\phi_0(T)]\):
    \begin{IEEEeqnarray*}{rClcl+rClcl}
        \phi_2^\eps &\xrightarrow[2]{\varphi}& \bar{\phi}_2 +[\phi_2]  &\quad\text{in} \quad& C([0,T]\times
        \mathcal{S}),&\theta_2^\eps&\xrightarrow[2]{\varphi}& \bar{\theta}_2 +[\theta_2] &\quad\text{in}
        \quad &C([0,T]\times \mathcal{S}),\\
        y_2^\eps &\xrightarrow[2]{\varphi}& \bar{y}_2+[y_2] &\quad\text{in} \quad &
        C([0,T]\times \mathcal{S}),& p_2^\eps &\xrightarrow[2]{\varphi}& \bar{p}_2+ [p_2] &\quad\text{in}
        \quad& C([0,T]\times \mathcal{S}),\\
        \IEEEeqnarraymulticol{10}{c}{
            \theta_1^\eps \xrightarrow[2]{\varphi} [\theta_1]  \quad\text{in}\quad  C([0,T]\times \mathcal{S}),
        }
    \end{IEEEeqnarray*}
    where \((\bar{\phi}_2, \bar{\theta}_2, \bar{y}_2, \bar{p}_2)\) is the unique solution to the initial value
    problem~\eqref{eq:38} with~\eqref{eq:a42} and where for \(t\in[0,T]\) and \(s\in \mathcal{S}\) we define
    \begin{IEEEeqnarray*}{rCl+rCl}
        [\theta_1](t, s) &\coloneqq&
        -\frac{\theta_\ast D_t L_0(t)}{2\omega(y_0(t))}\sin \left( 2\pi s \right),&[\phi_2](t, s) &\coloneqq&
        -\frac{D_t L_0(t)}{4\omega(y_0(t))}\cos(2\pi s),\\
        \phantom{}[y_2](t, s)&\coloneqq& -\frac{\theta_\ast  D_y L_0(t)}{4\omega(y_0(t))}\cos(2\pi s),&[p_2](t, s)
        &\coloneqq& \frac{d}{dt}\left( \frac{\theta_\ast  D_y L_0(t)}{4\omega(y_0(t))} \right)\cos(2\pi  s)
    \end{IEEEeqnarray*}
    and
    \begin{IEEEeqnarray*}{rCl}
        \phantom{}[\theta_2](t,s) &\coloneqq& -\theta_\ast D_y L_0(t) [y_2](t,s)
        - \frac{p_0(t)}{\omega(y_0(t))} [p_2](t,s)+ \frac{\theta_\ast^2 (D_y L_0(t))^2}{16 \omega(y_0(t))}\cos(4\pi s) \\
        && - \> \frac{\theta_\ast D_t L_0(t) }{\omega(y_0(t))}\bar{\phi}_2(t)\cos(2\pi s).
    \end{IEEEeqnarray*}
\end{proposition}

\begin{proof}
    The equivalence follows from Theorem~\ref{thm:2} and~\cite[Proposition 2.4]{Visintin2006b}.
\end{proof}

Note that in Proposition~\ref{cor:1} the constant \(\pi\) was chosen to normalise the period of the rapidly oscillating functions.

\subsection{Asymptotic expansion in the multidimensional case}

The asymptotic expansion results presented in this article can be generalised to the case of multiple fast and
slow DOF. This case was discussed in~\cite{Klar2021}, where the governing Lagrangian generalises by the natural extension of~\eqref{eq:03} to the case of multiple fast and slow DOF, i.e., \(y_\eps\in \RR^n\) and \(z_\eps\in \RR^r\) for \(n,r \geq 1\). In the case \(r>1\), it is possible that the \(z_\eps\) are in resonance with each other. After introducing some constraints on the system in form of two non-resonance conditions, a similar result to Theorem~\ref{thm:2} can be derived. These non-resonance conditions bypass the so-called ``small divisor problem'' which commonly emerges when deriving higher-order asymptotic expansions of trajectories of systems of multiple fast DOF (see~\cite[Chap.~7]{Sanders2007}). Non-resonance conditions also arise in connection with the KAM (Kolmogorov-Arnol'd-Moser) theorem~\cite{Arnold1983,Lazutkin1993} to similarly handle small divisors.

The proof is similar to the one presented in this
article. First, one introduces a change of coordinates \((z_\eps,\zeta_\eps)\mapsto (\theta_\eps, \phi_\eps)\)
via a generalised version of the generating function~\eqref{eq:a70} and one defines sequences of scaled
residual terms \(\{\theta_1^\eps\}, \{\phi_2^\eps\}, \{y_2^\eps\}, \{p_2^\eps\}\), and \(\{\theta_2^\eps\}\)
similar to Definition~\ref{def:1}. Next, one shows that the sequences \(\{\theta_1^\eps\}\) and
\(\{\phi_2^\eps\}\) are uniformly bounded in \(L^\infty([0,T],\RR^r)\) and \(\{y_2^{\eps}\}\) and
\(\{p_2^{\eps}\}\) are uniformly bounded in \(L^\infty([0,T], \RR^n)\). After deriving the oscillatory terms,
which is possible due to the transformation of the fast DOF into action-angle variables and
integration by parts, the weak\(^\ast\) limit can be derived with the extended Arzel\`a-Ascoli
Theorem~\cite[Principle~4, Chapter~I §1]{Bornemann1998a}, similar to the sketch of proof in Section~\ref{sec:2}. Since
\(\theta_2^\eps\in \RR^r\), a simple derivation from the energy, as described in this article for the case
\(r=1\), is not possible. Thus, more convoluted calculations are necessary to prove its uniform bound in
\(L^\infty([0,T], \RR^r)\), and its convergence in the weak\(^\ast\) limit.

\section{Thermodynamic expansion and interpretation}
\label{sec:6}

We now give a thermodynamic interpretation of the analytic result presented in Theorem~\ref{thm:2}. Thermodynamic effects can, in principle, occur when a separation of scales exist; Hertz
developed a thermodynamic theory for Hamiltonian systems which are perturbed by slow external agents~\cite{Hertz1910a}. The model considered in this article is an example of this kind, if we restrict
the analysis to the fast DOF (variable \(z_\eps\)) and consider the slow DOF
(variable \(y_\eps\)) as an external agent. The question we want to address in this section is ``Can we
replace the dynamics of the fast DOF with a thermodynamic description in terms of temperature
and entropy?'' As it turns out, even in this simple model problem, an interesting adiabatic/non-adiabatic
characteristic emerges through a higher-order asymptotic expansion.

The concept of adiabatic invariance finds applications in the analysis of slowly perturbed dynamical systems,
where one is primarily interested in the derivation of the effective evolution of the system. It emerged in
celestial mechanics in the form of the perturbation theory of Hamiltonian dynamical systems~\cite[Chapter 10]{Arnold1989a}, and can be found in many other fields~\cite{Li2019b, Evans2016, Duca2020, Neishtadt2019}. In particular, adiabatic invariance plays a crucial
role in thermodynamics. There, \emph{adiabatic thermodynamic processes} are idealised models in the limit of
an infinite separation of timescales and are, by definition, processes with constant entropy~\cite{Weiner2002a}.

For a thermodynamic interpretation of the system, we will regard the fast DOF \(z_\eps\) as the
system's thermal vibrations, acting on the slow DOF \(y_\eps\), which represents the system's
slow (mechanical) dynamics. As such, we will mainly focus the thermodynamic analysis on the energy
associated with the fast DOF \(E_\eps^\perp\), in contrast to the residual energy
\(E_\eps^\parallel\), which describes the remaining part of the system. Both energies can be read off from the
total energy~\eqref{eq:09}, i.e.,
\begin{equation}
    \label{eq:27}
    E_\eps^\perp = \tfrac{1}{2}\dot{z}_\eps^2+\tfrac{1}{2}\eps^{-2}\omega^2(y_\eps) z_\eps^2,
    \qquad E_\eps^\parallel= E_\eps - E_\eps^\perp.
\end{equation}

Note that the evolution of the fast DOF \(z_\eps\) is governed by the energy
\(E_\eps^\perp = E^\perp_\eps(z_\eps,\dot{z}_\eps;y_\eps)\), which is subject to a dynamically varying
external agent given by the evolution of the slow DOF \(y_\eps\). This framework allows us
to apply the theory developed by Hertz~\cite{Berdichevsky1997a}.

\subsection{The first and second law of thermodynamics}
\label{sec:1}

Hertz considered a thermodynamic system under a slowly changing external agent \(y\). A typical example is
a vessel filled with gas, where \(y\) indicates the height of a piston that compresses the gas. In line with
Hertz' analysis, we consider for the problem studied in this article the \emph{fast subsystem} with energy
\(E_\eps^\perp\) as a thermodynamic system under a slowly changing external agent represented by
\(y_\eps\). We point out that Hertz' theory is based on dynamical systems which are inherently reversible in
time and are thus regarded as idealised thermodynamic systems. As such, on a macroscopic level, the dynamics is
described by the first and second law of thermodynamics, where the second law is given in Carath\'eodory's
form~\cite{Weiner2002a}.

The first law states that every infinitesimal thermodynamic process, can be described as a change of the internal energy \(d A\), given by the work performed on the system via the application of an external force \(F\) along a displacement \(dy\), i.e., \(d A = F dy\), and the heat supply \(d Q\) such that the sum \(d A + d Q\) is the differential of some energy function \(E\),
\begin{equation*}
    d E = d A + d Q.
\end{equation*}

The second law of thermodynamics states that, for reversible processes, there are two functions of state,
namely the absolute \emph{temperature} \(T\) and the \emph{entropy} \(S\) such that
\begin{equation*}
    d Q = T d S.
\end{equation*}
Hence, the first and second law of thermodynamics combined read
\begin{equation}
    \label{eq:146}
    d E = F dy + T d S.
\end{equation}
Equation~\eqref{eq:146} can be reduced to the statement that there exists an entropy \(S=S(E,y)\) such that
the \emph{constitutive equations}
\begin{equation}
    \label{eq:142}
    \frac{1}{T} = \frac{\partial S(E,y)}{\partial E} \qquad \text{and} \qquad F = -T \frac{\partial S(E,y)}{\partial y}
\end{equation}
are satisfied.

The special case of a process without heat exchange, \(d Q = 0\), is called an \emph{adiabatic thermodynamic
    process}. In this case, all work \(d A\) is converted into a change of energy,
\begin{equation}
    \label{eq:147}
    d E = d A,
\end{equation}
or, equivalently, the entropy of the system stays constant,
\begin{equation*}
    S (E, y) = \mathrm{const}.
\end{equation*}

\begin{remark}
    With \(E=E(S,y)\), the constitutive equations~\eqref{eq:142} read equivalently
    \begin{equation}
        \label{eq:101}
        T=\frac{\partial E(S,y)}{\partial S} \qquad \text{and} \qquad F=\frac{\partial E(S,y)}{\partial y}.
    \end{equation}
\end{remark}

\subsection{Derivation of the thermodynamic quantities}

We follow for the derivation of the temperature, entropy, and external force the analysis in~\cite{Berdichevsky1997a}. This derivation was used, in a similar way, in~\cite{Klar2021} to determine the thermodynamic quantities
in the case of multiple fast DOF.

We consider the Hamiltonian and the associated energy of the fast subsystem given by
\begin{equation}
    \label{eq:last1}
    H_\eps^\perp(z_\eps, \zeta_\eps; y_\eps)= \frac{1}{2}\zeta_\eps^2 + \frac{1}{2} \eps^{-2} \omega^2(y_\eps) z_\eps^2 = E_\eps^\perp.
\end{equation}
Since \((z_\eps, \zeta_\eps)\) are fast relative to \(y_\eps\) we can, as a reasonable good approximation
assume that for one closed loop of \((z_\eps, \zeta_\eps)\) on the energy surface, the variable \(y_\eps\)
remains constant. This system is thus a harmonic oscillator, which is in one dimension ergodic on the energy surface (the
multidimensional case is discussed below). It thus admits a unique invariant measure, given by
\begin{equation*}
    \mu(A)= \frac{\displaystyle \int_{A} \dfrac{\ud \sigma}{\vert \nabla H^\perp_\eps \vert }}{\displaystyle \int_{\Sigma}
        \frac{\ud \sigma}
        {\vert \nabla H^\perp_\eps \vert }},
\end{equation*}

where \(A\subseteq \Sigma\) is a region on the level set
\(\Sigma= \{ (z_\eps, \zeta_\eps) \in \mathbb{R}^2\colon H^\perp_\eps(z_\eps, \zeta_\eps; y_\ast)= E^\perp_\ast \}\) and \(\ud \sigma\) is an infinitesimal small area element on the energy surface. In equilibrium thermodynamics, the temperature is considered a time independent variable, which is proportional to the time average of the kinetic energy of the system.

In general, if \(\left\langle \phi \right\rangle\) denotes the time average of some function \(\phi=\phi(t)\), the temperature \(T\) in equilibrium thermodynamics can be defined by the relation \(T= \left\langle 2 E_{\mathrm{kin}} \right\rangle\) where \(2 E_{\mathrm{kin}}=p \frac{\partial H(q, p)}{\partial p}\) and \((q,p)\) is a time-dependent trajectory on the energy surface in phase-space.

With the help of the ergodic theorem of Birkhoff and Khinchin (see, e.g.,~\cite{Walters1982a}), the invariant measure \(\mu\) can be used to derive the time average in the definition of the temperature by calculating the space average of \(p \frac{\partial H(q, p)}{\partial p}\) on the energy surface. In particular, in our example the temperature takes the form
\begin{equation}
    \label{eq:43}
    T_\eps(E^\perp_\ast, y_\ast)= \left\langle \zeta_\eps \frac{\partial H^\perp_\eps(z_\eps, \zeta_\eps; y_\ast)}{\partial \zeta_\eps} \right\rangle = \frac{\displaystyle \int_{\Sigma} \zeta_\eps \frac{\partial H^\perp_\eps}{\partial \zeta_\eps}\dfrac{\ud \sigma}{\vert \nabla H^\perp_\eps \vert }}
    {\displaystyle \int_{\Sigma} \frac{\ud \sigma}{\vert \nabla H^\perp_\eps \vert }}.
\end{equation}
Here, the numerator can be calculated using Gauss' theorem. It turns out that
\begin{equation}
    \label{eq:57}
    \int_{\Sigma} \zeta_\eps \frac{\partial H^\perp_\eps}{\partial \zeta_\eps}\dfrac{\ud \sigma}{\vert \nabla H^\perp_\eps \vert} = \Gamma_\eps(E^\perp_\ast, y_\ast),
\end{equation}
where \(\Gamma_\eps(E^\perp_\ast, y_\ast)\) is the phase-space volume enclosed by the trajectory of
\((z_\eps, \zeta_\eps)\) and is also known as the action of the orbit in Hamiltonian
dynamics~\cite{Goldstein2013a}.

To derive the denominator in~\eqref{eq:43}, we calculate the derivative of
\(\Gamma_\eps(E^\perp_\ast, y_\ast)\) with respect to \(E^\perp_\ast\) and find, with some geometrical
considerations, that
\begin{equation}
    \label{eq:44}
    \int_\Sigma\frac{\ud \sigma}{\vert \nabla H^\perp_\eps \vert} = \frac{\partial \Gamma_\eps(E^\perp_\ast,
        y_\ast)}{\partial E^\perp_\ast}.
\end{equation}
Thus, in equation~\eqref{eq:43} the numerator is given by~\eqref{eq:57} and the denominator by~\eqref{eq:44}, which allows us to express the temperature in terms of the phase-space volume \(\Gamma_\eps(E^\perp_\ast,y_\ast)\), i.e., \begin{equation}
    \label{eq:143}
    T_\eps(E^\perp_\ast, y_\ast) = \frac{\Gamma_\eps(E^\perp_\ast,y_\ast)}{\partial \Gamma_\eps(E^\perp_\ast,y_\ast)/\partial E^\perp_\ast}.
\end{equation}
According to the left equation in~\eqref{eq:142} we integrate \(T_\eps^{-1}\) in~\eqref{eq:143} with respect to \(E^\perp_\ast\) and
obtain as formula for the entropy
\( S_\eps(E^\perp_\ast, y_\ast) = \log \left( \Gamma_\eps(E^\perp_\ast, y_\ast) \right) + f_\eps(y_\ast)\).  It
remains to show that the function \(f_\eps\) is constant. To see this, we investigate the dependence of \(S_\eps\)
on \(y_\ast\) and will, similar to above, use~\eqref{eq:142}.

We can define the external force as the time-average of \(\partial_y H(q,p;y)
\) for a Hamiltonian system depending on some external agent \(y\) and apply, similar to above, the Birkhoff--Kinchin Theorem, to derive in our example
\begin{equation}
    \label{eq:45}
    F_\eps(E^\perp_\ast, y_\ast) = \left\langle \frac{\partial H^\perp_\eps(z_\eps, \zeta_\eps; y_\ast)}{\partial y_\ast}
    \right\rangle
    = \frac{\displaystyle \int_{\Sigma} \frac{\partial H^\perp_\eps}{\partial y_\ast}\dfrac{\ud \sigma}
        {\vert \nabla H^\perp_\eps \vert }}
    {\displaystyle \int_{\Sigma} \frac{\ud \sigma}{\vert \nabla H^\perp_\eps \vert }}.
\end{equation}

For the numerator, we calculate this time the derivative of \(\Gamma_\eps(E^\perp_\ast, y_\ast)\) with respect
to \(y_\ast\).

Similar to before, geometric considerations imply that
\begin{equation}
    \label{eq:48}
    \int_\Sigma \frac{\partial H^\perp_\eps}{\partial y_\ast}
    \frac{\ud \sigma}{\vert \nabla H^\perp_\eps \vert } = -\frac{\partial \Gamma_\eps(E^\perp_\ast, y_\ast)}{\partial y_\ast}.
\end{equation}
Combining equations~\eqref{eq:44},~\eqref{eq:45}, and~\eqref{eq:48} we obtain
\begin{equation}
    \label{eq:53}
    F_\eps(E^\perp_\ast, y_\ast) = -\frac{\partial \Gamma_\eps(E^\perp_\ast, y_\ast)/\partial y_\ast}{\partial
        \Gamma_\eps(E^\perp_\ast, y_\ast)/\partial E^\perp_\ast}.
\end{equation}
Once again, according to the right equation in~\eqref{eq:142}, we integrate \(-F_\eps/T_\eps\) in~\eqref{eq:143} and~\eqref{eq:53} with respect to \(y_\ast\) and find the desired formula for the entropy,
\begin{equation}
    \label{eq:54}
    S_\eps(E^\perp_\ast, y_\ast) = \log \left( \Gamma_\eps(E^\perp_\ast, y_\ast) \right) + C_\eps,
\end{equation}
where \(C_\eps\) is an arbitrary constant. This is a key result of Hertz: the entropy of a Hamiltonian system under
the influence of a slowly varying agent is, up to a constant, the logarithm of the phase-space volume.

The phase-space volume of a harmonic oscillator in one dimension with fixed energy \(E_\ast^\perp\) and fixed external agent \(y_\ast\) is geometrically described by an ellipse, thus using~\eqref{eq:105} and~\eqref{eq:last1} with \(E_\eps^\perp\) and \(y_\eps\) fixed yields
\begin{equation*}
    \Gamma_\eps(E^\perp_\ast, y_\ast) = 2\pi \eps \frac{E^\perp_\ast}{\omega(y_\ast)}= 2\pi \eps \theta_\ast.
\end{equation*}

Since on the fast timescale, slowly varying \(E^\perp_\eps\) and \(y_\eps\) can be considered to good approximation as constant, we argue by similarity and replace in equation~\eqref{eq:143},~\eqref{eq:53}, and~\eqref{eq:54} \(y_\ast\) by \(y_\eps\) and \(E_\ast^\perp\) by \(E_\eps^\perp\) so that
\begin{equation*}
    \Gamma_\eps(E^\perp_\eps, y_\eps) = 2\pi \eps \frac{E^\perp_\eps}{\omega(y_\eps)}= 2\pi \eps \theta_\eps.
\end{equation*}
To avoid a divergent entropy in the limit \(\eps\to0\), the constant in the entropy has to be chosen accordingly, for example \(C_\eps=-\log(\Gamma_\eps(E_\ast^\perp, y_\ast))\). In this case we would have
\begin{equation*}
    S_\eps(E_\eps^\perp, y_\eps) = \log \left( \frac{\Gamma_\eps(E^\perp_\eps, y_\eps)}{\Gamma_\eps(E_\ast^\perp, y_\ast)} \right) = \log \left(\frac{\theta_\eps}{\theta_\ast}\right) = \log ( \theta_\eps ) + C,
\end{equation*}
where \(C=-\log(\theta_\ast)\).

Thus, altogether, we derive for \(\eps>0\) the following expressions for the \emph{temperature} \(T_\eps\),
the \emph{entropy} \(S_\eps\), and the \emph{external force} \(F_\eps\) in the fast subsystem:
\begin{equation}
    \label{eq:49}
    T_\eps = \theta_\eps \omega(y_\eps),\qquad S_\eps =\log(\theta_\eps)+C,\qquad F_\eps = \theta_\eps \omega'(y_\eps).
\end{equation}

\subsection{Expansion of the thermodynamic quantities}

In combination with the second-order expansion derived in Theorem~\ref{thm:2} we will analyse the asymptotic
properties of the thermodynamic expressions above, by expanding \(y_\eps\) and \(\theta_\eps\)
in~\eqref{eq:49}, which in turn defines higher-order asymptotic expansions of the form
\(T_\eps= T_0 +T_1^\eps\), \(F_\eps= F_0+F_1^\eps\) and
\(S_\eps= S_0 +\eps [\bar{S}_1]^\eps +\eps^2 [\bar{S}_2]^\eps + \eps^2 S_3^\eps\) with \(T_1^\eps, F_1^\eps,S_3^\eps \to 0\) in
\(C([0,T])\), where
\begin{equation}
    \label{eq:35a}
    T_0\coloneqq \theta_\ast \omega(y_0), \qquad F_0 \coloneqq \theta_\ast \omega'(y_0),
\end{equation}
and
\begin{equation}
    \label{eq:35b}
    S_0 \coloneqq \log(\theta_\ast)+C, \qquad  [\bar{S}_1]^\eps \coloneqq \frac{[\theta_1]^\eps}{\theta_\ast},
    \qquad [\bar{S}_2]^\eps
    \coloneqq \frac{\bar{\theta}_2 +[\theta_2]^\eps}{\theta_\ast}
    -\frac{1}{2}\left( \frac{[\theta_1]^\eps}{\theta_\ast} \right)^2.
\end{equation}
Analogously, by substituting~\eqref{eq:105} into~\eqref{eq:27}, we expand the energy of the fast subsystem
\(E^\perp_\eps=\theta_\eps\omega(y_\eps)=E^\perp_0 + \eps [\bar{E}_1^\perp]^\eps+ \eps^2 [\bar{E}_2^\perp]^\eps +\eps^2 E_3^{\perp\eps}\) where \(E_3^{\perp\eps} \to 0\) in \(C([0,T])\) and obtain
\begin{equation}
    \label{eq:35c}
    E_0^\perp \coloneqq \theta_\ast \omega(y_0),\qquad [\bar{E}_1^\perp]^\eps\coloneqq \omega(y_0)
    \left[ \theta_1 \right]^\eps,
\end{equation}
\begin{equation}
    \label{eq:a67}
    [\bar{E}_2^\perp]^\eps
    \coloneqq \theta_\ast
    \omega'(y_0)\left( \bar{y}_2+\left[ y_2 \right]^\eps \right)+ \omega(y_0)\left( \bar{\theta}_2
    + \left[\theta_2 \right]^\eps \right).
\end{equation}

\subsection{Leading-order thermodynamics}

We show in this section that to leading-order (limit \(\eps\to0\)), the resulting thermodynamic process is
adiabatic or, in other words, a thermodynamic process with constant entropy.

In fact, by comparing this result with the leading-order expansion of the temperature, entropy, and external
force in the model problem as derived in~\eqref{eq:35a} and~\eqref{eq:35b}, i.e.,
\begin{equation*}
    T_0= \theta_\ast\omega(y_0),\qquad S_0 = \log(\theta_\ast)+C, \qquad F_0=\theta_\ast\omega'(y_0),
\end{equation*}
we observe that \(S_0\equiv\mathrm{const.}\) and hence reason that the limit process can be interpreted as an
adiabatic thermodynamic process. This is in particular a result of the external agent \(y_0\), which
affects the fast system to leading-order only slowly. The energy in the limit \(\eps \to 0\) (see
equation~\eqref{eq:35c}) is given by \(E^\perp_0(y_0) = \theta_\ast \omega(y_0)\), and therefore (cf.\
equation~\eqref{eq:147})
\begin{equation}
    \label{eq:E-decom-adia}
    d E^\perp_0 = F_0 d y_0 +T_0 d S_0 = F_0 d y_0,
\end{equation}
which resembles equation~\eqref{eq:146}.
\begin{remark}
    \label{remark:1}
    Note that by action and reaction, the force exerted on the fast subsystem \(F_0\) is equal but of opposite
    sign to the force acting on the slow DOF, i.e., \(\ddot{y}_0=-F_0\) (see also~\eqref{eq:4}).
\end{remark}

\subsection{Second-order thermodynamics}

We calculate the average contribution of the higher-order microscale oscillations in \(E^\perp_\eps\) and
\(S_\eps\) by taking the weak\(^\ast\) limit of the asymptotic expansion terms \([\bar{E}_1^\perp]^\eps\),
\([\bar{E}_2^\perp]^\eps\), \([\bar{S}_1]^\eps\), and \([\bar{S}_2]^\eps\) in~\eqref{eq:35b}--\eqref{eq:a67}. With property~\eqref{eq:56}, i.e., \([\theta_1]^\eps, [y_2]^\eps, [\theta_2]^\eps \weak{\ast} 0\) in \(L^\infty([0,T])\) this yields
\begin{equation*}
    [\bar{E}_1^\perp]^\eps\weak{\ast}0 \quad \text{in}\quad L^\infty([0,T]),\qquad [\bar{S}_1]^\eps \weak{\ast} 0 \quad \text{in} \quad L^\infty([0,T]),
\end{equation*}
and
\begin{IEEEeqnarray}{rClCl}
    [\bar{E}_2^\perp]^\eps
    &\weak{\ast}&\bar{E}_2^\perp\coloneqq\theta_\ast \omega'(y_0) \bar{y}_2+ \omega(y_0)\bar{\theta}_2&\quad \text{in}
    \quad& L^\infty([0,T]), \label{eq:wlim-E2}\\
    \phantom{}[\bar{S}_2]^\eps &\weak{\ast}&\bar{S}_2\coloneqq \frac{\bar{\theta}_2}{\theta_\ast} - \left( \frac{D_t L_0}{4 \omega(y_0)} \right)^2& \quad \text{in} \quad& L^\infty([0,T]). \label{eq:wlim-S2}
\end{IEEEeqnarray}

We can now define the second-order average energy and entropy
\begin{equation*}
    \bar{E}^\perp_\eps \coloneqq E^\perp_0 + \eps^2 \bar{E}^\perp_2,\qquad \bar{S}_\eps \coloneqq S_0+ \eps^2 \bar{S}_2.
\end{equation*}

To analyse the thermodynamic properties of \(\bar{E}_\eps^\perp\) for \(\eps>0\), we focus on the second-order
expansion term \(\bar{E}_2^\perp\). The appropriate temperature, entropy, and external force can be read off from~\eqref{eq:wlim-E2} and~\eqref{eq:wlim-S2}:
\begin{equation*}
    T_0= \theta_\ast\omega(y_0),\qquad \bar{S}_2 = \frac{\bar{\theta}_2}{\theta_\ast}
    -\left( \frac{D_t L_0}{4 \omega(y_0)} \right)^2,\qquad F_0=\theta_\ast\omega'(y_0).
\end{equation*}

Note that the entropy in this case is \emph{not} constant. This can be explained by the second-order
correction of the external agent \(y_\eps\), which exhibits according to Theorem~\ref{thm:2} a
decomposition into a slowly varying component \(\bar{y}_2\) and a rapidly varying component
\(\left[ y_2 \right]^\eps\). By considering
\(\bar{E}^\perp_2 = \bar{E}^\perp_2(\bar{S}_2,\bar{y}_2; y_0, p_0)\) we obtain (cf.\ equation~\eqref{eq:146})
for fixed \((y_0, p_0)\)
\begin{equation*}
    d \bar{E}^\perp_2 = F_0 d \bar{y}_2 + T_0 d \bar{S}_2,
\end{equation*}
which agrees with the qualitative discussion of Section~\ref{sec:1}.

Finally, let us inspect how the thermodynamic energy balance is realised in the total second-order energy
correction of \(E_\eps\). Analogous to the decomposition~\eqref{eq:27}, we split the total energy \(E_\eps\)
in~\eqref{eq:26} into \(E_\eps^\perp\) and \(E_\eps^\parallel\), i.e.,
\(E_\eps =E^\parallel_\eps + E^\perp_\eps\), where
\begin{equation*}
    E^\parallel_\eps = \frac{1}{2}p_\eps^2+\frac{\eps}{2}\theta_\eps p_\eps  D_y L_\eps \sin(2\eps^{-1}\phi_\eps)
    +\frac{\eps^2}{8}\theta_\eps^2 (D_y L_\eps)^2\sin^2(2\eps^{-1}\phi_\eps).
\end{equation*}
We then use the expressions derived in Theorem~\ref{thm:2} to expand
\(E^\parallel_\eps=E^\parallel_0 + \eps [\bar{E}_1^\parallel]^\eps + \eps^2 [\bar{E}_2^\parallel]^\eps +\eps^2 E_3^{\parallel\eps}\), where \(E_3^{\parallel\eps}\to0\) in \(C([0,T])\) with
\begin{equation*}
    E_0^\parallel \coloneqq\frac{1}{2}p_0^2,\qquad [\bar{E}_1^\parallel]^\eps\coloneqq\frac{\theta_\ast}{2}
    D_t L_0\sin(2\eps^{-1}\phi_0),
\end{equation*}
and
\begin{IEEEeqnarray}{rCl}
    \label{eq:a68}
    [\bar{E}_2^\parallel]^\eps&\coloneqq&  p_0 \left( \bar{p}_2+\left[ p_2 \right]^\eps \right) +\frac{\theta_\ast^2 (D_y L_0)^2}{8}\sin^2(2\eps^{-1}\phi_0)\\
    && + \theta_\ast D_t L_0\left( \bar{\phi}_2+[\phi_2]^\eps \right)\cos(2\eps^{-1}\phi_0)+ \frac{ [\theta_1]^\eps D_t L_0}{2}\sin (2\eps^{-1}\phi_0). \IEEEnonumber
\end{IEEEeqnarray}

As before, we take the weak\(^\ast\) limit to determine the average energy correction at first- and
second-order, and find
\begin{IEEEeqnarray*}{rCl}
    [\bar{E}_1^\parallel]^\eps&\weak{\ast}&0 \quad \text{in}\quad L^\infty([0,T]), \\
    \phantom{}[\bar{E}_2^\parallel]^\eps
    &\weak{\ast}&\bar{E}_2^{\parallel}\coloneqq p_0 \bar{p}_2 +\left( \frac{\theta_\ast D_y L_0}{4} \right)^2
    -\frac{\theta_\ast (D_t L_0)^2}{4\omega(y_0)} \quad \text{in}\quad L^\infty([0,T]),
\end{IEEEeqnarray*}
and define the averaged residual energy
\(\bar{E}^\parallel_\eps \coloneqq E_0^\parallel+ \eps^2 \bar{E}_2^\parallel\). The following theorem shows
how the Hamiltonian character of the model problem and the thermodynamic interpretation materialise for the
averaged total second-order energy correction \(\bar{E}_2 = \bar{E}^\parallel_2 + \bar{E}^\perp_2\).

\begin{theorem}
    \label{thm:3}
    Let \((y_0, p_0)\) be as in~\eqref{eq:59} and \((\bar{y}_2, \bar{p}_2)\) be as in Theorem~\ref{thm:2}. Let
    \(\bar{E}_2\) be the averaged total second-order energy correction
    \(\bar{E}_2= \bar{E}_2^\parallel + \bar{E}_2^\perp\), where
    \begin{equation*}
        \bar{E}_2^\parallel(\bar{y}_2, \bar{p}_2; y_0, p_0) = p_0 \bar{p}_2 +\frac{\theta_\ast^2 \omega'(y_0)^2}{16 \omega^2(y_0)}
        -\frac{\theta_\ast (p_0 \omega'(y_0))^2}{4\omega^3(y_0)},
    \end{equation*}
    and
    \begin{equation*}
        \bar{E}_2^\perp(\bar{y}_2; y_0, p_0) = \theta_\ast \omega'(y_0) \bar{y}_2+ \omega(y_0)\bar{\theta}_2(y_0, p_0),
    \end{equation*}
    with
    \begin{equation*}
        \bar{\theta}_2(y_0, p_0) = \frac{\theta_\ast p_0 (\omega'(y_0))^2}{8 \omega^4(y_0)} + C_{\theta}, \qquad C_{\theta}= -\frac{\theta_\ast p_\ast (\omega'(y_\ast))^2}{8 \omega^4(y_\ast)} -[\theta_2]^\eps(0).
    \end{equation*}
    Then the differential equation~\eqref{eq:40} and~\eqref{eq:41} take the form
    \begin{equation*}
        \frac{d  \bar{y}_2 }{dt} = \frac{\partial \bar{E}_2}{\partial p_0},\qquad \frac{\partial \bar{p}_2}{dt}
        = -\frac{\partial \bar{E}_2 }{\partial y_0}.
    \end{equation*}
    Additionally, with the functions \(T_0\), \(\bar{S}_2\), and \(F_0\), which can be interpreted as the temperature, entropy, and external force in the fast subsystem, the energy \(\bar{E}_2^\perp\) can be written as
    \begin{equation*}
        \bar{E}_2^\perp(\bar{S}_2, \bar{y}_2; y_0, p_0)= F_0 \bar{y}_2 + T_0 \bar{S}_2 + \frac{\theta_\ast (p_0 \omega'(y_0))^2}{16 \omega^3(y_0)}.
    \end{equation*}
    With this notation, the energy \(\bar{E}_2^\perp\) satisfies the constituent equations~\eqref{eq:101} in the form
    \begin{equation*}
        F_0 = \frac{\partial \bar{E}_2}{\partial \bar{y}_2}, \qquad T_0 = \frac{\partial \bar{E}_2}{\partial \bar{S}_2}.
    \end{equation*}
\end{theorem}

\begin{proof}
    The claim follows directly from~\eqref{eq:40} and~\eqref{eq:41}.
\end{proof}

\subsection{Thermodynamic interpretation in the multidimensional case}
The discussion in~\cite{Klar2021} shows how to derive the second-order asymptotic expansion in the case of
multiple fast and slow DOF. Similar to the expansion covered in this article, one can interpret
the dynamics to leading-order as well as to second-order from a thermodynamic point of view. If there is only
one fast DOF, the fast subsystem is ergodic and thus Hertz' theory can be applied directly. In particular (see Remark~\ref{remark:1}) the derived force to leading-order of the
internal energy~\eqref{eq:E-decom-adia} coincides (up to sign) with the force of the slow DOF
derived through the weak\(^\ast\) limit~\eqref{eq:4}. If there are more than one fast DOF,
however, the fast subsystem is in general not ergodic. Thus, Hertz' theory has to be adapted in
equations~\eqref{eq:43} and~\eqref{eq:45} by considering the ensemble average instead of the time
average. This allows to derive expressions which correspond to temperature, entropy, and external force. These
expressions can be expanded to second order, similar to the procedure described in this article. It turns out
that in this case, energy relations appear which likewise resemble the first and second law of
thermodynamics. However, in contrast to the case of one fast DOF, the derived force to leading-order of the internal energy does not coincide (up to sign) with the force of the slow degrees of freedom derived through weak\(^\ast\) convergence techniques.

While the force derived in the latter case corresponds to the total force the slow DOF experience, the force derived in the former case is due only to a change of the internal energy. In fact, if we assume that temperature is kept constant in the system or if we consider only a very small time-frame such that \(y_0\approx y_\ast\), then the force derived from the internal energy coincides with the force derived from the Fixman-potential assuming equi-distribution of energy in the normal vibrational modes (see~\cite{Bornemann1998a}). Thus, similar to~\cite{Bornemann1998a}, the Fixman-potential provides an inexact description of the slow dynamics of the system. An exact description of the slow dynamics can be obtained by considering further contributions from the mutual interactions with the fast DOF. This information can be obtained from the entropic term.

\section*{Conclusions}
In this article, we discussed a simple fast-slow mechanical system. It represents a minimal example of a class
of related models. The expansion to second order and its thermodynamic interpretation carry over to finitely many
DOF. It is a natural next step to consider the extension to real-world applications in
molecular dynamics. In molecular dynamics one is usually interested in the slow macroscopic dynamics; the fast
microscopic dynamics is less important but is necessary to obtain an accurate representation of the dynamics
on the microscopic scale. Through averaging techniques, it is possible (see for instance~\cite{Bornemann1998a}) to derive a homogenised system, which describes the dynamics only of the slow
DOF. However, this homogenised system has to be understood as an approximation to the slow
dynamics of the original system because it is derived by considering the limit \(\eps\to0\). For a more
detailed description of the slow dynamics in this system, a higher-order asymptotic expansion is of interest,
as the scale-parameter \(\eps\) is a quotient of mass ratios and thus in many applications small but finite.

The fast-slow mechanical system studied in this article is one of the simplest models that can potentially
exhibit thermodynamic effects. Indeed, one of the core assumptions of thermodynamics is that the system under
consideration has a clear separation of scales. This assumption is by construction satisfied for our
system. Many theories have been developed to study large-scale interacting particle systems from a
thermodynamic point of view, but only a few focus on applying the theory in combination with the averaging
methods for dynamical systems.

One of the earliest attempts to describe fast-slow mechanical systems from a thermodynamic point of view dates
back to the work of Hertz~\cite{Hertz1910a}. Since the fast subsystem in our model is ergodic, Hertz'
methodology can directly be applied to the fast-slow system studied in this article. 
We derived the second-order asymptotic expansion of the dynamics in the system via weak convergence techniques and expressed in addition the results in the framework of two-scale
convergence. Moreover, we showed that the dynamics to leading-order as well as on average to second-order
satisfy equations resembling the first and second law of thermodynamics. In the limit \(\eps\to0\), the
expression of the internal energy can be used to derive the same evolution equation for the slow dynamics as
by methods based on homogenisation procedures~\cite{Bornemann1998a}. For \(\eps>0\), i.e., for the average
second-order asymptotic expansion, a similar statement cannot be made because the entropic term does not
vanish, implying more complicated dynamics to second-order, which cannot just be described by the internal
energy.

From the first and second law of thermodynamics, it is clear that a simple description of the slow dynamics
can be achieved when the entropic term vanishes, which is the case if the fast dynamics is ergodic. In the
case of multiple fast DOF, the fast dynamics is, in general, not ergodic. Thus, a direct
application of Hertz' theory is not possible. As demonstrated in~\cite{Klar2021} however, some adjusted
techniques can be used to derive a thermodynamic interpretation of the system. It turns out that the leading-order and the averaged second-order dynamics do not have a vanishing ergodic term and thus the internal energy
alone cannot be used to describe the slow dynamics in the system. Contributions from the entropic term need to
be considered to describe the slow dynamics in the system correctly.

The discussion in this article provides two different perspectives on the asymptotic expansion of solutions to our model problem~\eqref{eq:04}. Firstly, the rigorous derivation of the asymptotic expansion of the system's dynamics provides accurate information about the evolution of the system up to second order. Secondly, the thermodynamic interpretation up to second order provides a bridge to the realm of statistical mechanics. A possible application of the theory presented in this article is to approximate the dynamics of system~\eqref{eq:04} by a system that describes the slow evolution accurately, but the fast evolution by thermodynamic expressions in terms of temperature and entropy. Whether the evolution of these quantities can be derived is an open problem, as is the role of entropy for thermodynamic consistency. Yet, the asymptotic expansion can be used profitable in numerical simulations~\cite{Klar2021}.

\subsubsection*{Acknowledgements}
MK was supported by a scholarship from the EPSRC Centre for Doctoral Training in Statistical Applied Mathematics at Bath (SAMBa), under the project EP/L015684/1. JZ gratefully acknowledges funding by a Royal Society Wolfson Research Merit Award. All authors thank Celia Reina for stimulating discussions and helpful suggestions throughout this project.


\end{document}